\documentclass[a4paper,11pt]{article}
\pdfoutput=1 

\usepackage{jcappub} 

\usepackage[T1]{fontenc} 

\usepackage{bm,latexsym,amsmath,amssymb,amsfonts,mathrsfs}
\allowdisplaybreaks[1]
\newcommand*{\D}{\mathrm{d}}

\newcommand*{\dpi}{\delta\pi_{\ell m}}
\newcommand*{\tdPhi}{\widetilde{\delta\Phi}_{\ell m}}
\newcommand*{\tPhi}{\widetilde{\Phi}}
\newcommand*{\TK}{\textcolor[rgb]{1.0,0.1,0.1}}

\title{Unscreened multipole moments of the fifth force in the EFT of dark energy}

\author{Tsutomu~Kobayashi}
\author{and Toshiki~Takadera}
\affiliation{Department of Physics, Rikkyo University, Toshima, Tokyo 171-8501, Japan}

\emailAdd{tsutomu@rikkyo.ac.jp}
\emailAdd{t\_takadera@rikkyo.ac.jp}

\abstract{It has been argued that degenerate higher-order scalar-tensor theories and the effective field theory (EFT) of dark energy are endowed with the Vainshtein mechanism, resulting in a screened fifth force in the exterior of a gravitational source.
However, spherical symmetry has been assumed in most of the discussions so far.
In this paper, we study whether the Vainshtein mechanism operates in the EFT of dark energy beyond spherical symmetry, focusing in particular on the role of the ``beyond Horndeski'' EFT parameters.
Assuming that deviations from spherical symmetry are small, we compute multipole moments of the gravitational potential for a given nonspherical source.
For a generic choice of the ``beyond Horndeski'' EFT parameters, it is shown that the multipole moments of the fifth force are not screened in the region where the monopole component is screened.
Rather, the gravitational potential shows a characteristic oscillatory behavior in its multipole components.
In the special case where the EFT parameters are tuned so that graviton decay into dark energy is practically absent, the behavior of the multipole moments is qualitatively different from that in the generic case.
However, also in this case, the Vainshtein mechanism is not efficient enough to screen the fifth force around the source.}

\begin{document}
\maketitle
\flushbottom

\section{Introduction}\label{sec:intro}

Accounting for the current accelerated expansion of the Universe has long been one of the central aims of cosmology.
Recent observational data show a mild preference for an evolving dark energy equation of state~\cite{DESI:2024mwx, DESI:2025zgx, DES:2025sig, DES:2026jmi}.
This situation motivates us to explore modified gravity as a possible underlying theory beyond $\Lambda$CDM.
Among various theories with different gravitational degrees of freedom, scalar-tensor theories offer the simplest but sufficiently effective framework for studying aspects of modified gravity.
Since the rediscovery~\cite{Charmousis:2011bf, Deffayet:2011gz, Kobayashi:2011nu} of the Horndeski theory~\cite{Horndeski:1974wa}, efforts have been made to extend the theory space of healthy scalar-tensor theories.
As the Horndeski theory is the most general scalar-tensor theory with second-order field equations, its generalization requires higher-order field equations, which would result in the appearance of unstable Ostrogradsky modes~\cite{Ostrogradsky:1850fid, Woodard:2015zca}.
A way out from this difficulty is to consider degenerate higher-order scalar-tensor (DHOST) theories~\cite{Langlois:2015cwa, Crisostomi:2016czh, BenAchour:2016cay}.
A part of the space of DHOST theories can be mapped from the Horndeski theory~\cite{Zumalacarregui:2013pma} via disfromal transformations~\cite{Bekenstein:1992pj}.
See refs.~\cite{Langlois:2018dxi,Kobayashi:2019hrl} for reviews.
Although the degeneracy of the kinetic matrix is imposed in any coordinate system in the original construction of DHOST theories, this requirement can be relaxed, and one may just impose the degeneracy in the particular gauge in which the scalar field serves as the time coordinate~\cite{DeFelice:2018ewo}.
Recently, this idea was combined with the higher-derivative generalization of disformal transformations~\cite{Takahashi:2021ttd, Takahashi:2022mew} to systematically formulate healthy scalar-tensor theories with third-order derivatives of the scalar field in the action~\cite{Michiwaki:2026xru} (see also ref.~\cite{Gorji:2026nes} for a similar attempt).
The entire model space of these scalar-tensor theories is extremely large because the action depends on a number of free functions of the scalar field and its derivatives.

An efficient way of confronting such a broad class of scalar-tensor theories with cosmological observations is the use of the effective field theory (EFT) of dark energy and modified gravity~\cite{Gubitosi:2012hu, Bloomfield:2012ff, Gleyzes:2013ooa, Gleyzes:2014rba}, encompassing all models in the Horndeski/DHOST family (and possibly further generalizations with a single scalar degree of freedom).
The EFT is obtained directly by expanding the Horndeski/DHOST action in terms of perturbations to relevant order around a cosmological background.
The EFT action is characterized by several time-dependent functions, which are expressed as specific combinations of the free functions in the original action evaluated at the background~\cite{Bellini:2014fua}.
These EFT functions determine the behavior of the perturbations.
See ref.~\cite{Frusciante:2019xia} for a review of the EFT of dark energy and modified gravity.

If one modifies general relativity on cosmological scales by adding an extra degree of freedom, the modification could persist down to small scales, leading to inconsistencies with solar system experiments and astrophysical tests.
It is therefore required that the fifth force mediated by the extra field be screened within a certain distance from a source.
Proposed screening mechanisms include chameleon screening~\cite{Khoury:2003aq}, symmetron screening~\cite{Hinterbichler:2010es}, Vainshtein screening~\cite{Vainshtein:1972sx}, and k-mouflage~\cite{Babichev:2009ee}, each of which relies on a different kind of nonlinearities of the scalar field.
Among these, the Vainshtein screening mechanism manifests itself in galileon-type scalar field theories~\cite{Nicolis:2008in} due to their nonlinear derivative self-interactions.
Given that the Horndeski theory~\cite{Horndeski:1974wa} is equivalently formulated as the generalized galileon theory~\cite{Deffayet:2011gz, Kobayashi:2011nu}, it is natural to expect that the Horndeski/DHOST family of scalar-tensor theories and their EFT description are endowed with the Vainshtein mechanism.\footnote{Of course, the Horndeski theory can accommodate chameleon screening as well. See ref.~\cite{Sirera:2026klo} for a recent attempt at a unified framework for different screening mechanisms.}
It has been shown that the Vainshtein mechanism indeed suppresses the fifth force in the Horndeski theory~\cite{Kimura:2011dc, Narikawa:2013pjr, Koyama:2013paa}.
Screening in scalar-tensor theories beyond Horndeski has a richer phenomenological structure: Vainshtein screening is partially broken inside material bodies.
This was first noticed in ref.~\cite{Kobayashi:2014ida} within the Gleyzes--Langlois--Piazza--Vernizzi generalization of the Horndeski theory~\cite{Gleyzes:2014dya}, and was further discussed in the context of generic DHOST theories in refs.~\cite{Dima:2017pwp, Langlois:2017dyl, Crisostomi:2017lbg}.
It was also found that the effective Newton constant is different inside and outside material bodies~\cite{Hirano:2019scf, Crisostomi:2019yfo} in the special subset of the DHOST family that prohibits gravitons from decaying into the scalar field~\cite{Creminelli:2018xsv}.

It should be noted, however, that the previous works mentioned above have examined Vainshtein screening for static and spherically symmetric configurations.
For cylindrical and planar sources, the profile of a flat-space galileon coupled to matter
was investigated analytically, leading to the conclusion that the suppression of the galileon-mediated force is sensitive to the shape of the source (or, more precisely, the dimensionality of the system)~\cite{Bloomfield:2014zfa}.
The implications of this shape-dependence for the cosmic large-scale structure have been discussed in refs.~\cite{Falck:2014jwa, Falck:2015rsa, Burrage:2019afs}.
Even in the case of a cubic galileon in flat space, sophisticated numerical techniques are required for dealing with the inherent derivative nonlinearity in less symmetric systems such as a two-body system~\cite{Hiramatsu:2012xj, Ogawa:2018srw, White:2020xsq}.
Incorporating the time dependence, scalar gravitational radiation from binary systems in flat-space galileon theories has been elaborated in refs.~\cite{deRham:2012fw, deRham:2012fg, Dar:2018dra, deRham:2024xxb}.
Screening in nonspherical systems in DHOST theories remains far less explored and understood:
axisymmetric solutions for slowly rotating stars were studied in ref.~\cite{Anson:2020fum}, and the parity-odd tidal response of relativistic stars was investigated in ref.~\cite{Kobayashi:2025bdh}.

The purpose of the present paper is to push forward the understanding of Vainshtein screening beyond spherical symmetry in DHOST theories and the EFT of dark energy.
Using the effective Lagrangian obtained under the static approximation, we study the gravitational fields produced by non-spherical sources.
To avoid solving nonlinear partial differential equations, we assume that deviations from spherical symmetry are small and can be treated as perturbations around a spherical configuration.
The profile of a flat-space galileon around a deformed body was obtained in this way in ref.~\cite{Brito:2014ifa}.
A similar perturbation scheme was employed for flat-space galileon theories to compute the Green's function and the scalar-mediated force in refs.~\cite{Chu:2012kz, Andrews:2013qva}.
The present work extends ref.~\cite{Brito:2014ifa} to include various terms characterizing DHOST theories and the EFT of dark energy.

This paper is organized as follows.
In the next section, we review the effective Lagrangian for the Vainshtein mechanism derived from the EFT of dark energy and DHOST theories, and present a spherically symmetric solution to see how Vainshtein screening and its partial breaking occur.
Small deviations from spherical symmetry are considered in section~\ref{sec:multip}.
We demonstrate that the multipole moments of the fifth force are not screened, and the gravitational potential shows characteristic large deviations from the standard results in its multipole components.
In section~\ref{sec:special-case}, we investigate  the special case where the parameters of the EFT of dark energy satisfy a certain relation and Vainshtein screening operates in a different way from the generic case.
We draw our conclusions in section~\ref{sec:disc}.

\section{Vainshtein screening in the EFT of dark energy}\label{sec:eft-of-de}

\subsection{Effective Lagrangian}

Let us work in the EFT of dark energy/scalar-tensor theories with a luminal gravitational-wave speed.
We consider static scalar perturbations in the Newtonian gauge on scales where the cosmic expansion can be ignored, and write the fluctuation of the scalar field as $\pi(\Vec{x})$ and the metric as
\begin{align}
    \D s^2=-\left(1+2\Phi\right)\D t^2+\left(1-2\Psi\right)
    \D\Vec{x}^2,\label{metric:newtonian-gauge}
\end{align}
Expanding the EFT Lagrangian in terms of the perturbations, we obtain the $\alpha$-basis~\cite{Bellini:2014fua, Langlois:2017mxy} effective Lagrangian~\cite{Dima:2017pwp} (see also~\cite{Langlois:2017dyl, Crisostomi:2017lbg} and appendix~\ref{app:eft-derivation})
\begin{align}
        \mathcal{L}_{\textrm{eff}}&=\frac{M^2}{2}\biggl\{
                \left(
                        \eta_0\pi+\eta_1\Phi+\eta_2\Psi
                \right)\nabla^2\pi + 4(1+\alpha_H)\Psi\nabla^2\Phi -2 \Psi\nabla^2\Psi 
                -\beta_3\Phi\nabla^2\Phi 
                \notag \\ & \quad 
                +r_c^2\eta_3\mathcal{L}_3^{\textrm{Gal}}
                -2r_c^2(\alpha_H+2\beta_1) \Phi\mathcal{E}_3^{\textrm{Gal}}
                +r_c^2\left[-4\alpha_H\Psi_i+2(2\beta_1+\beta_3)\Phi_i \right]
                \mathcal{X}_i
                \notag \\ & \quad 
                +
                        2r_c^4(\alpha_H+2\beta_1)\mathcal{L}_4^{\textrm{Gal}}
                        +r_c^4(4\beta_1+\beta_3)
                        \mathcal{X}_i\mathcal{X}_i
        \biggr\}-\Phi \rho,
        \label{effLag}
\end{align}
where
\begin{align}
        &\mathcal{E}_3^{\textrm{Gal}}:=(\nabla^2\pi)^2-\pi_{ij}\pi_{ij},
        \qquad 
        \mathcal{X}:=\frac{1}{2}\pi_i\pi_i,
        \notag \\ &
        \mathcal{L}_3^{\textrm{Gal}}:=-\mathcal{X}\nabla^2\pi,
        \qquad 
        \mathcal{L}_4^{\textrm{Gal}}:=-\mathcal{X}\mathcal{E}_3^{\textrm{Gal}}.
\end{align}
and we adopted the notation $\pi_i:=\nabla_i\pi$, $\pi_{ij}:=\nabla_{i}\nabla_{j}\pi$, etc.
Here we introduced the length scale $r_c$ so that the field $\pi$ is dimensionless.
The parameters $\eta_0, \eta_1, \eta_2, \eta_3, \alpha_H, \beta_1$, and $\beta_3$ are also dimensionless.
These parameters depend on time, but it is natural to assume that they vary on cosmological time scales.
We thus ignore the time dependence and treat these parameters as constants.
The degeneracy condition implies $\beta_3=-2\beta_1[2(1+\alpha_H)+\beta_1]$~\cite{Langlois:2017mxy}.
The last term in eq.~\eqref{effLag} represents the coupling with nonrelativistic matter with the energy density $\rho(\Vec{x})$.
In what follows, we absorb the factor $M^2(1-\alpha_H-3\beta_1)$ into $\rho$ and write $\rho/M^2(1-\alpha_H-3\beta_1)\to \rho$ so that the effective Newton constant is equal to $1/8\pi$ (as will become clear below).

Among the EFT parameters in the Lagrangian, we are interested in the impacts of $\alpha_H$ and $\beta_1$.
These two parameters characterize genuine ``beyond Horndeski'' effects, and setting them to zero leaves us with a subset of the Horndeski family with a luminal speed of gravitational waves.
This strongly motivates us to elucidate gravitational phenomena specific to dark energy/scalar-tensor models with nonvanishing $\alpha_H$ and $\beta_1$.

The equations of motion derived from the Lagrangian~\eqref{effLag} are
\begin{align}
    &\frac{\eta_1}{2}\nabla^2\pi+2(1+\alpha_H)\nabla^2\Psi-\beta_3\nabla^2\Phi
    -r_c^2(\alpha_H+2\beta_1)\mathcal{E}_3^{\textrm{Gal}}
    -r_c^2(2\beta_1+\beta_3)\nabla^2\mathcal{X}
    \notag \\ &
    =(1-\alpha_H-3\beta_1)\rho,
    \label{eq:g-Phi}
    \\ 
    &\frac{\eta_2}{2}\nabla^2\pi+2(1+\alpha_H)\nabla^2\Phi-2\nabla^2\Psi+2r_c^2\alpha_H
    \nabla^2\mathcal{X}=0,
    \label{eq:g-Psi}
    \\ 
    &\eta_0\nabla^2\pi+\frac{\eta_1}{2}\nabla^2\Phi+\frac{\eta_2}{2}\nabla^2\Psi
    +\frac{r_c^2\eta_3}{2}\mathcal{E}_3^{\textrm{Gal}}
    -2r_c^2(\alpha_H+2\beta_1)\left[\nabla^2\Phi\nabla^2\pi
    -\Phi_{ij}\pi_{ij}\right]
    \notag \\ &
    +r_c^2\nabla_i\left\{\left[-2\alpha_H\nabla^2\Psi+(2\beta_1+\beta_3)\nabla^2\Phi\right]
    \pi_i\right\}
    +r_c^4(\alpha_H+2\beta_1)\mathcal{E}_4^{\textrm{Gal}}
    \notag \\ &
    +r_c^4(4\beta_1+\beta_3)\nabla_i\left(
        \pi_i\nabla^2\mathcal{X}
    \right)=0,
    \label{eq:g-pi}
\end{align}
where
\begin{align}
    \mathcal{E}_4^{\textrm{Gal}}:=(\nabla^2\pi)^3-3\pi_{ij}\pi_{ij}\nabla^2\pi
    +2\pi_{ij}\pi_{jk}\pi_{ki}.
\end{align}
To remove the higher derivatives acting on $\pi$, it is convenient to introduce
\begin{align}
    \tPhi&:=\Phi+\frac{\alpha_H+\beta_1}{1+\alpha_H+\beta_1}r_c^2\mathcal{X},
    \label{def:tildePhi}
    \\
    \widetilde\Psi &:=\Psi+\frac{\beta_1}{1+\alpha_H+\beta_1}r_c^2\mathcal{X}
    \label{def:tildePsi}.
\end{align}
From eqs.~\eqref{eq:g-Phi} and~\eqref{eq:g-Psi}, we obtain
\begin{align}
    \nabla^2\tPhi&=
    \frac{\xi}{2}\nabla^2\pi+
    \frac{1-\alpha_H-3\beta_1}{2(1+\alpha_H+\beta_1)^2}\rho
    +\frac{\alpha_H+2\beta_1}{2(1+\alpha_H+\beta_1)^2}r_c^2\mathcal{E}_3^{\textrm{Gal}},
    \label{eq:tildePhi}
    \\
    \nabla^2\widetilde\Psi&=
    \frac{-(1+\alpha_H)\eta_1+\beta_1(2+2\alpha_H+\beta_1)\eta_2}{4(1+\alpha_H+\beta_1)^2}\nabla^2\pi
    +\frac{(1+\alpha_H)(1-\alpha_H-3\beta_1)}{2(1+\alpha_H+\beta_1)^2}\rho
    \notag \\ &\quad 
    +\frac{(1+\alpha_H)(\alpha_H+2\beta_1)}{2(1+\alpha_H+\beta_1)^2}r_c^2\mathcal{E}_3^{\textrm{Gal}},
    \label{eq:tildePsi}
\end{align}
where
\begin{align}
    \xi:=-\frac{\eta_1+(1+\alpha_H)\eta_2}{2(1+\alpha_H+\beta_1)^2}.
\end{align}
Using eqs.~\eqref{eq:tildePhi} and~\eqref{eq:tildePsi}, one can rewrite the scalar field equation~\eqref{eq:g-pi} as
\begin{align}
    &\hat\eta_0\nabla^2\pi
    -\frac{\xi}{2}(1-\alpha_H-3\beta_1)
    \rho+\frac{r_c^2\hat\eta_3}{2}
    \mathcal{E}_3^{\textrm{Gal}}
    -2r_c^2(\alpha_H+2\beta_1)\left[\nabla^2\tPhi
    \nabla^2\pi-\tPhi_{ij}\pi_{ij}\right]
        \notag \\ &
    -r_c^2\frac{(\alpha_H+\beta_1)(1-\alpha_H-3\beta_1)}{1+\alpha_H+\beta_1}
    \nabla_i\left(
        \pi_i \rho
    \right)+r_c^4\left(\frac{\alpha_H+2\beta_1}{1+\alpha_H+\beta_1}\right)
    \mathcal{E}_4^{\textrm{Gal}}
    =0,
    \label{eq:reduce-scalar-eq}
\end{align}
with
\begin{align}
    \hat\eta_0&:=\eta_0
    -\frac{(\eta_1-\beta_1\eta_2)[\eta_1+(2+2\alpha_H+\beta_1)\eta_2]}%
    {8(1+\alpha_H+\beta_1)^2}
    \\
    \hat\eta_3&:=
    \eta_3-\xi(\alpha_H+2\beta_1)+\frac{\eta_1\alpha_H+(\eta_1+\eta_2)\beta_1}{1+\alpha_H+\beta_1}.
\end{align}
Equations~\eqref{eq:tildePhi} and~\eqref{eq:reduce-scalar-eq} provide a coupled system for $\tPhi$ and $\pi$.
To determine $\Psi$, one can use eq.~\eqref{eq:tildePsi}, or instead integrate eq.~\eqref{eq:g-Psi} and use
\begin{align}
    \Psi=(1+\alpha_H)\Phi+\frac{\eta_2}{4}\pi+r_c^2\alpha_H\mathcal{X}.
    \label{relation-Phi-Psi}
\end{align}

We are now in a position to compare the structure of the basic equations derived from the EFT of dark energy with that of a flat-space galileon.
As summarized briefly in appendix~\ref{app:flat-space-galileon}, in the case of a flat-space galileon we solve the nonlinear equation of motion for $\pi$ in the presence of the matter source $\rho$, and then compare the fifth force from the galileon, $\nabla_i\pi$, with the standard Newtonian force $\nabla_i\Phi_{\textrm{N}}$ derived from the Poisson equation $\nabla^2\Phi_{\textrm{N}}=\rho/2$.
In the case of the EFT of dark energy, we note the following differences as compared to the flat-space galileon.
First, the equation of motion for $\pi$ depends on the derivative of the source through $\nabla_i(\pi_i\rho)$ in eq.~\eqref{eq:reduce-scalar-eq}.
This causes the partial breaking of Vainshtein screening inside material bodies~\cite{Kobayashi:2014ida, Koyama:2015oma, Saito:2015fza}.
Second, the gravitational potential manifests itself in the equation of motion for $\pi$.
In the case of spherical symmetry, one can use eq.~\eqref{eq:tildePhi} to remove the gravitational potential from the equation of motion for $\pi$.
This procedure then brings a new source-dependent term in the equation of motion for $\pi$, which plays a crucial role in determining the screened profile of $\pi$.
Note that the terms of the form $\nabla\nabla\tPhi\nabla\nabla\pi$ in eq.~\eqref{eq:reduce-scalar-eq} cannot be removed by further field redefinitions analogous to eqs.~\eqref{def:tildePhi} and~\eqref{def:tildePsi}.
Third, the actual gravitational potential is determined through eqs.~\eqref{def:tildePhi} and~\eqref{eq:tildePhi}, which depend nonlinearly on the derivatives of $\pi$.
The resultant gravitational potential $\Phi$ (or its gradient) is compared with the standard Newtonian potential $\Phi_{\textrm{N}}$ to see whether and how the fifth force is suppressed.

\subsection{Spherically symmetric solution}

Let us derive a spherically symmetric solution to see how the Vainshtein mechanism operates in the EFT of dark energy.
Here and in the next section we assume that $\alpha_H+2\beta_1\neq 0$.
We need a separate treatment for the special case of $\alpha_H+2\beta_1=0$, which is discussed later in section~\ref{sec:special-case}.

We label the spherically symmetric solution with a subscript 0, so that $\tPhi=\tPhi_0(r)$, $\pi=\pi_0(r)$, and $\rho=\rho_0(r)$, where $r$ is the radial coordinate.
It is convenient to introduce $\mu(r)$ defined by
\begin{align}
    \mu'(r)=r^2\rho_0(r),
\end{align}
where a prime denotes differentiation with respect to $r$.
We also define the following dimensionless quantities:
\begin{align}
    x_0(r):=\frac{r_c^2\pi_0'}{r},\qquad \tilde y_0(r):=\frac{r_c^2\tPhi_0'}{r},
    \qquad 
    \mathcal{A}(r):=\frac{r_c^2\mu}{r^3}.
\end{align}
Equations~\eqref{eq:tildePhi} and~\eqref{eq:reduce-scalar-eq} can be integrated once, leading respectively to
\begin{align}
    \tilde y_0=\frac{\xi}{2}x_0
    +\frac{1-\alpha_H-3\beta_1}{2(1+\alpha_H+\beta_1)^2}\mathcal{A}
    +\frac{\alpha_H+2\beta_1}{(1+\alpha_H+\beta_1)^2}x_0^2,
    \label{eq:y0}
\end{align}
and
\begin{align}
    &\hat\eta_0x_0-\frac{\xi}{2}(1-\alpha_H-3\beta_1)
    \mathcal{A}+\hat\eta_3x_0^2
    -4\left(\alpha_H+2\beta_1\right)x_0 \tilde y_0
    \notag \\ &
    -\frac{(\alpha_H+\beta_1)(1-\alpha_H-3\beta_1)}{1+\alpha_H+\beta_1}
    \frac{\left(r^3\mathcal{A}\right)'}{r^2}x_0
    +2\left(\frac{\alpha_H+2\beta_1}{1+\alpha_H+\beta_1}\right)x_0^3=0.
    \label{eq:x0-y0}
\end{align}
One can use eq.~\eqref{eq:y0} to remove $\tilde y_0$ from eq.~\eqref{eq:x0-y0} to obtain a cubic polynomial equation for $x_0$.

If $\mathcal{A}\ll 1$, one may linearize eqs.~\eqref{eq:y0} and~\eqref{eq:x0-y0} to obtain
\begin{align}
    x_0&=
    \frac{\xi}{2\hat\eta_0}(1-\alpha_H-3\beta_1)
    \mathcal{A},
    \label{lin-sol0-1}
    \\
    \tilde y_0&=\left[\frac{\xi^2}{2\hat\eta_0}
    +\frac{1}{2(1+\alpha_H+\beta_1)^2}\right](1-\alpha_H-3\beta_1)\mathcal{A}.
    \label{lin-sol0-2}
\end{align}
In this linear regime, one has $\Phi_0\neq\Psi_0$ for a generic choice of the parameters.

In the regime where $\mathcal{A}\gg 1$ and hence the terms nonlinear in $x_0$ are important, we have
\begin{align}
    x_0^3-\mathcal{A}x_0-\frac{(1+\alpha_H+\beta_1)(\alpha_H+\beta_1)}{2(\alpha_H+2\beta_1)}
    \frac{\left(r^3\mathcal{A}\right)'}{r^2}x_0\simeq
    0.
\end{align}
The solution in this regime is thus given by
\begin{align}
    x_0^2&\simeq \mathcal{A}+\frac{(1+\alpha_H+\beta_1)(\alpha_H+\beta_1)}{2(\alpha_H+2\beta_1)}
    \frac{\left(r^3\mathcal{A}\right)'}{r^2},
    \label{soln-x0}
    \\ 
    \tilde y_0&\simeq \frac{1}{2(1+\alpha_H+\beta_1)}\left[
        \mathcal{A}+(\alpha_H+\beta_1)\frac{\left(r^3\mathcal{A}\right)'}{r^2}
    \right]
    \label{soln-y0},
\end{align}
where note that $\tilde y_0=\mathcal{O}(x_0^2)\gg 1$.
More explicitly, we have
\begin{align}
    \pi_0'&\simeq \frac{\sigma}{r_c}\sqrt{
    \frac{\mu}{r}+\frac{(1+\alpha_H+\beta_1)(\alpha_H+\beta_1)}{2(\alpha_H+2\beta_1)}\mu'
    },
    \label{soln:pi0}
    \\
    \tPhi_0'&\simeq
     \frac{1}{2(1+\alpha_H+\beta_1)}\left[
        \frac{\mu}{r^2}+(\alpha_H+\beta_1)\frac{\mu'}{r}
    \right],
\end{align}
where $\sigma=\pm 1$.
Note that the sign $\sigma$ is not determined if one inspects the solution only locally in the nonlinear regime.
However, one of the signs is selected by requiring that the solution is connected to the appropriate one in the linear regime~\eqref{lin-sol0-1}.
Using eq.~\eqref{def:tildePhi}, we obtain
\begin{align}
    \Phi_0'\simeq\frac{1}{2}
    \left[
        \frac{\mu}{r^2}-\frac{(\alpha_H+\beta_1)^2}{2(\alpha_H+2\beta_1)}
        \mu''
    \right].
    \label{eq:spherical-Phi}
\end{align}
We then use eq.~\eqref{relation-Phi-Psi} to derive
\begin{align}
    \Psi_0'\simeq\frac{1}{2}
    \left[
        \frac{\mu}{r^2}+\alpha_H\frac{\mu'}{r}
        -\frac{\beta_1(\alpha_H+\beta_1)}{2(\alpha_H+2\beta_1)}
        \mu''
    \right].
    \label{eq:spherical-Psi}
\end{align}
The terms dependent on $\mu'$ and $\mu''$ signal the partial breaking of Vainshtein screening~\cite{Kobayashi:2014ida, Koyama:2015oma, Saito:2015fza}, which is activated only inside material bodies.
In the external region where $\rho_0=0$, we have $\mu=\mu_0=\;$const, and hence the two potentials coincide: $\Phi_0'=\Psi_0'=\mu_0/2r^2$.
Thus, it is convenient to introduce the fiducial Newtonian potential $\Phi_{\textrm{N}}$ derived from
\begin{align}
    \nabla^2\Phi_{\textrm{N}}=\frac{\rho}{2},
\end{align}
which is compared with $\Phi$ to see how efficiently the Vainshtein mechanism operates.
In the exterior of a spherical source and in the nonlinear regime, we have $\Phi_0=\Psi_0=\Phi_{\textrm{N}0}$.

    \begin{figure}[tb]
        \begin{center}
                \includegraphics[keepaspectratio=true,height=60mm]{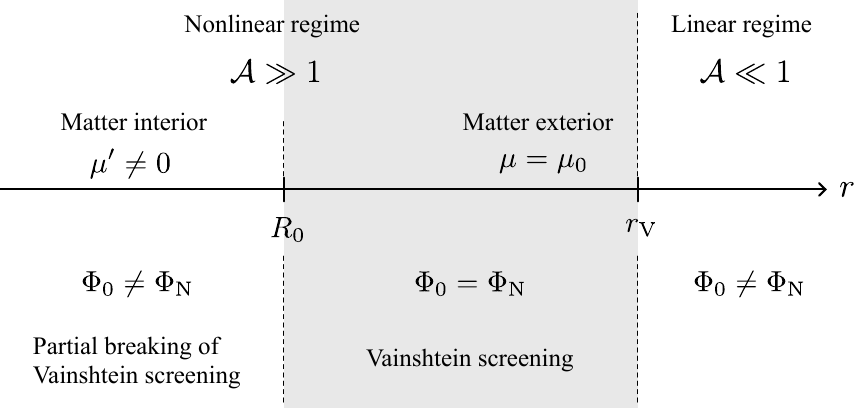}
        \end{center}
                \caption{Vainshtein screening in the case of spherical symmetry.}
        \label{fig:regimes.pdf}
    \end{figure}

Let us define the Vainshtein radius as $r_{\textrm{V}}:=\left(r_c^2\mu_0\right)^{1/3}$, so that $\mathcal{A}\gg 1$ for $r\ll r_{\textrm{V}}$.
(We assume that the location satisfying $\mathcal{A}=1$ is in the external region.)
We have three different regions with different behavior of the gravitational potential and the fifth force: (i) $0\le r\lesssim R_0$, (ii) $R_0\lesssim r\lesssim r_{\textrm{V}}$, and (iii) $r_{\textrm{V}}\lesssim r$, where $R_0$ is the size of the source.
The situation is summarized in figure~\ref{fig:regimes.pdf}.

\begin{figure}
  \begin{minipage}{0.48\columnwidth}
    \centering
    \includegraphics[width=\linewidth]{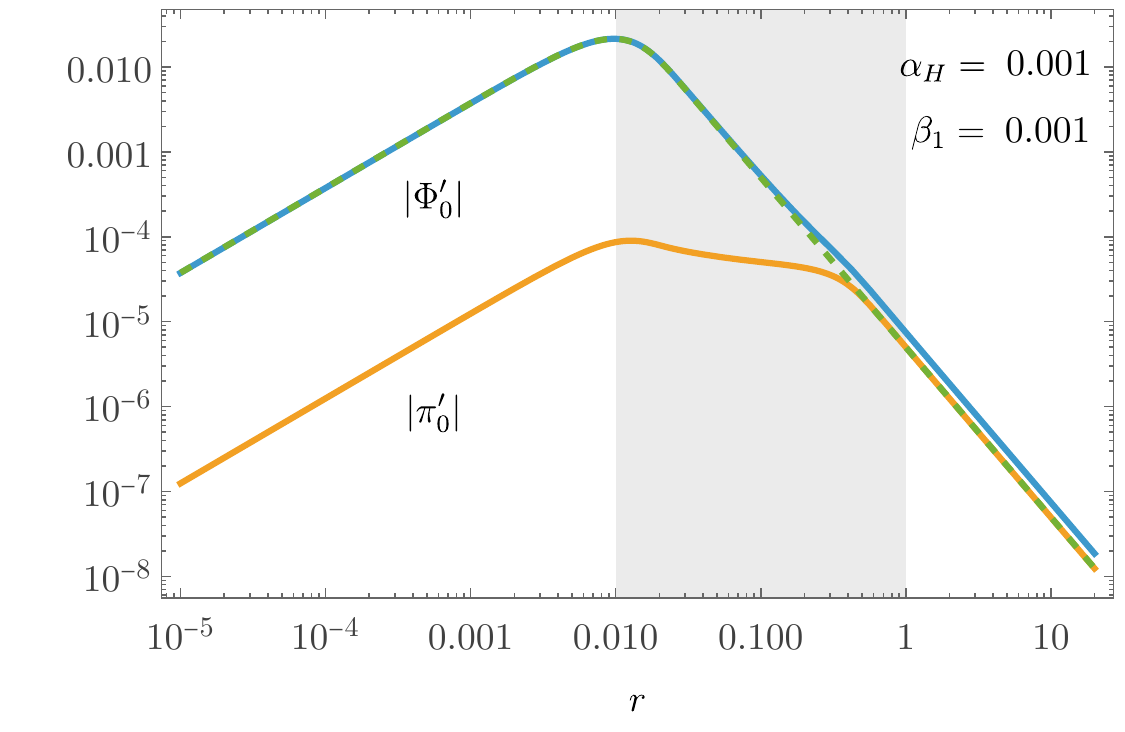}
  \end{minipage}
  \hfill
  \begin{minipage}{0.48\columnwidth}
    \centering
    \includegraphics[width=\linewidth]{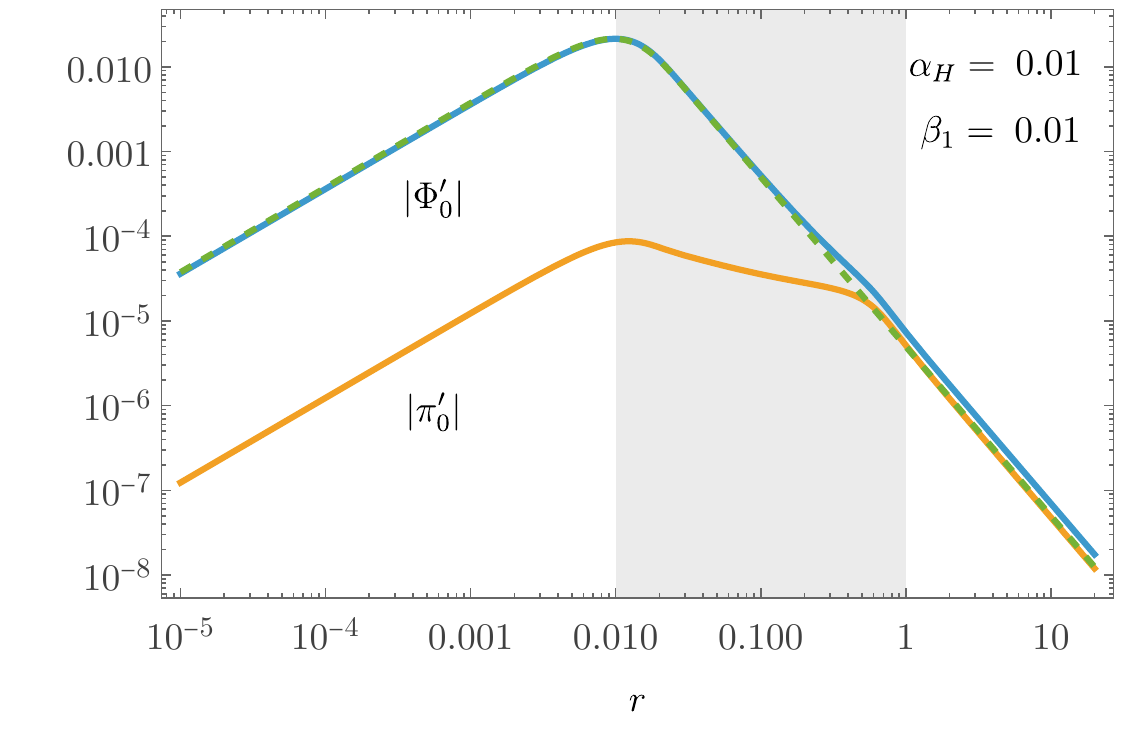}
  \end{minipage}
  \caption{The radial gradient of the gravitational potential (blue solid lines) is compared with the standard Newtonian result, $\Phi_{\textrm{N}}'=\mu(r)/2r^2$ (green dashed lines).
  The radial gradient of the scalar field is also shown (orange solid lines).
  The EFT parameters are given by $\hat\eta_0=1$, $\xi=1$, and $\hat\eta_3=0$ for both plots, while $\alpha_H$ and $\beta_1$ are chosen as shown in each plot.
  The parameters characterizing the density are chosen as $R_0=10^{-2}r_{\textrm{V}}$ and $\mu_0=10^{-3}R_0$ with $r_{\textrm{V}}=1$.
  The shaded region corresponds to $R_0\le r\le r_{\textrm{V}}$.}
  \label{fig:L=0}
\end{figure}

As an example of spherically symmetric configurations, let us consider the Gaussian profile,
\begin{align}
    \rho_0(r)&=\frac{4\mu_0}{\sqrt{\pi}R_0^3}e^{-r^2/R_0^2},
    \label{gauss-density-0}
    \\ 
    \mu(r)&=\mu_0\left[
        \textrm{erf}(r/R_0)-\frac{2re^{-r^2/R_0^2}}{\sqrt{\pi}R_0}
    \right],
\end{align}
where erf$(x)$ is the error function.
This profile does not have a definite surface, but $R_0$ can be regarded as the characteristic size of the source, as $\mu(r)\simeq \mu_0$ for $r\gtrsim 2R_0$.
For selected values of the parameters, the solution to eqs.~\eqref{eq:y0} and~\eqref{eq:x0-y0} is presented in figure~\ref{fig:L=0}, down to the linear regime where the Vainshtein mechanism no longer works.
It can be seen that the fifth force is suppressed quite well for $r\lesssim 0.1 \times r_{\textrm{V}} \ll r_{\textrm{V}}$.
Since we take $\alpha_H,\beta_1\ll 1$, only a tiny breaking of Vainshtein screening is seen to occur for $r\lesssim R_0$.

Before closing this section, let us give the following short comment.
For the above screened solution to exist, the expression in the square root in eq.~\eqref{soln:pi0} must be positive.
In particular, we have $\mu\simeq \rho_c r^3/3$ in the vicinity of the center, where $\rho_c$ is the central density, yielding the constraint on the parameters,
\begin{align}
    c_0:=1+\frac{3(\alpha_H+\beta_1)(1+\alpha_H+\beta_1)}{2(\alpha_H+2\beta_1)}\ge 0.
    \label{def:c0}
\end{align}
We consider the parameters satisfying this constraint.

\section{Multipole moments of the gravitational potential}\label{sec:multip}

\subsection{Small deviations from spherical symmetry}

By carefully following the calculations in the previous section, one notices that the recovery of standard gravity in the region $R_0\lesssim r\ll r_{\textrm{V}}$ is a consequence of a delicate interplay among different terms, which casts doubt on the validity of the conclusion beyond spherical symmetry.
We are thus motivated to study the solutions to eqs.~\eqref{eq:tildePhi} and~\eqref{eq:reduce-scalar-eq} in the case of a nonspherical source.\footnote{We present the analysis of the case with cylindrical symmetry in appendix~\ref{app:cylindrical}, for which the multipole expansion employed in this section is not directly applicable.}
We assume that the source, and hence the scalar field and the gravitational potential, deviate only slightly from spherical symmetry:
\begin{align}
    \rho&=\rho_0(r)+\delta\rho(r,\theta,\varphi),
    \\
    \tPhi&=\tPhi_{0}(r)+ \widetilde{\delta\Phi}(r,\theta,\varphi),
    \label{eq:expand-tPhi}
    \\
    \pi&=\pi_0(r)+\delta\pi(r,\theta,\varphi),
\end{align}
where the spherical coordinate system $(r,\theta,\varphi)$ is used.
The zeroth-order spherically symmetric part is determined by solving eqs.~\eqref{eq:y0} and~\eqref{eq:x0-y0}.
To describe the small deviations from spherical symmetry, it is useful to employ a multipole expansion and decompose $\delta\rho$, $\delta\tPhi$, and $\delta\pi$ using the spherical harmonics $Y_{\ell m}(\theta,\varphi)$ as
\begin{align}
    \delta\rho&=\sum_{\ell m}\delta\rho_{\ell m} (r)Y_{\ell m}(\theta,\varphi),\label{m-ex:rho}
    \\
    \widetilde{\delta\Phi}&=\sum_{\ell m}\tdPhi(r)
    Y_{\ell m}(\theta,\varphi),
    \\
    \delta\pi&=\sum_{\ell m}\dpi(r)Y_{\ell m}(\theta,\varphi).\label{m-ex:pi}
\end{align}

In the case of standard Newtonian gravity with the Poisson equation $\nabla^2\delta\Phi_{\textrm{N}}=\delta\rho/2$, such a multipole expansion gives the ordinary differential equation
\begin{align}
    \frac{1}{r^2}\left(r^2\delta\Phi'_{\textrm{N}\ell m}\right)'-\frac{\ell(\ell+1)}{r^2}\delta\Phi_{\textrm{N}\ell m}=
    \frac{\delta\rho_{\ell m}}{2}.
\end{align}
The Green's function can be constructed using the two homogeneous solutions, $r^\ell$ and $r^{-\ell-1}$, and then $\delta\Phi_{\ell m}$ can be obtained for any $\delta\rho_{\ell m}$.
In the exterior region where $\delta\rho_{\ell m}=0$, the solution is simply given by
\begin{align}
    \delta\Phi_{\textrm{N}\ell m}&=-\frac{1}{2(2\ell+1)}\frac{I_{\ell m}}{r^{\ell+1}},
    \label{eq:newton-result}
    \\ 
    I_{\ell m}&:=\int \delta\rho_{\ell m}(r)r^{\ell+2}\D r.
\end{align}
Here, $I_{\ell m}$ is the multipole moment of the mass distribution, with the domain of integration being the volume occupied by the matter.

In the case of the EFT of dark energy, the substitution of eqs.~\eqref{m-ex:rho}--\eqref{m-ex:pi} into eqs.~\eqref{eq:tildePhi} and~\eqref{eq:reduce-scalar-eq} yield, to first order in perturbations, the coupled ordinary differential equations for $\tdPhi(r)$ and $\dpi(r)$,
\begin{align}
    &\left[
        \frac{1}{r^2}\left(r^2\tdPhi'\right)'-\frac{\ell(\ell+1)}{r^2}\tdPhi
    \right] -\frac{\xi}{2}
    \left[
        \frac{1}{r^2}\left(r^2\dpi'\right)'-\frac{\ell(\ell+1)}{r^2}\dpi
    \right]
    -\frac{(1-\alpha_H-3\beta_1)}{2(1+\alpha_H+\beta_1)^2}\delta\rho_{\ell m} 
    \notag \\ &=
    \frac{\alpha_H+2\beta_1}{(1+\alpha_H+\beta_1)^2}\left[
        \frac{2}{r^2}\left(r^2x_0\dpi'\right)'
    -\frac{\ell(\ell+1)}{r^3}\left(r^2x_0\right)'\dpi
    \right], 
    \label{pert-eq-1}
\end{align}
and
\begin{align}
    &\left(\frac{1+\alpha_H+\beta_1}{\alpha_H+2\beta_1}\right)\biggl\{
        \hat\eta_0\left[
        \frac{1}{r^2}\left(r^2\dpi'\right)'-\frac{\ell(\ell+1)}{r^2}\dpi
        \right]
        -\frac{\xi}{2}(1-\alpha_H-3\beta_1)\delta\rho_{\ell m}
        \notag \\ & \quad
        +\hat \eta_3
        \left[\frac{2}{r^2}\left(r^2x_0\dpi'\right)'
            -\frac{\ell(\ell+1)}{r^3}\left(r^2x_0\right)'\dpi\right]
    \biggr\}
    \notag \\ 
    &-2(1+\alpha_H+\beta_1)\biggl[
        \frac{2}{r^2}\left(r^2x_0\tdPhi'\right)'-\ell(\ell+1)
    \frac{(r^2x_0)'}{r^3}\tdPhi
    \notag \\ &
    +\frac{2}{r^2}\left(r^2\tilde y_0\dpi'\right)'-\ell(\ell+1)
    \frac{(r^2\tilde y_0)'}{r^3}\dpi
    \biggr]
    \notag \\ & 
    -\frac{(\alpha_H+\beta_1)(1-\alpha_H-3\beta_1)}{\alpha_H+2\beta_1}
    \biggl[
        \frac{1}{r^2}\left(r^3x_0\delta\rho_{\ell m}\right)'
    +\frac{r_c^2}{r^2}\left(\mu'\dpi'\right)'-\frac{\ell(\ell+1)r_c^2\mu'}{r^4}\dpi
    \biggr]
    \notag \\ &
    +3\left[
        \frac{2}{r^2}\left(
        r^2x_0^2\dpi'
    \right)'-\frac{\ell(\ell+1)}{r^3}\left(r^2x_0^2\right)'\dpi
    \right]=0.
    \label{pert-eq-2}
\end{align}
Given the source term $\delta\rho_{\ell m}(r)$, we integrate eqs.~\eqref{pert-eq-1} and~\eqref{pert-eq-2}, subject to appropriate boundary conditions at $r=0$ and $r\to \infty$.
The multipole moments of the gravitational potential are then obtained from the relation that follows from the linearization of eq.~\eqref{def:tildePhi}:
\begin{align}
    \delta\Phi_{\ell m}
    =    \tdPhi-r_c^2
    \left(\frac{\alpha_H+\beta_1}{1+\alpha_H+\beta_1}\right)\pi_0'\dpi'.
    \label{rel:tildephi-phi}
\end{align}
One can also decompose the deviation $\delta\Psi$ of $\Psi$ from spherical symmetry as $\delta\Psi=\sum\delta\Psi_{\ell m}Y_{\ell m}$.
Then, from eq.~\eqref{relation-Phi-Psi} we obtain
\begin{align}
    \delta\Psi_{\ell m}=(1+\alpha_H)\delta\Phi_{\ell m}
    +r_c^2\alpha_H\pi_0'\dpi'+\frac{\eta_2}{4}\dpi.
\end{align}

To assess whether the Vainshtein mechanism efficiently screens the multipole moments of the fifth force within the Vainshtein radius in the EFT of dark energy, we proceed as follows.
First, we investigate whether or not $\delta\Phi_{\ell m}$ reproduces the expected $\sim r^{-\ell-1}$ behavior inside the Vainshtein radius and in the exterior of the source.
Of course, this is not sufficient, and we then check whether or not the two potentials coincide, $\delta\Phi_{\ell m}=\delta\Psi_{\ell m} (=\delta\Phi_{\textrm{N}\ell m})$.

Suppose that the zeroth-order spherically symmetric solution is in the linear regime, $r\gg r_{\textrm{V}}$. 
We then have $x_0\ll 1$ and $\tilde y_0\ll 1$.
Let us first derive the solution in this linear regime.
Dropping $\delta\rho_{\ell m}$ and all the small terms that depend on $x_0$ and $\tilde y_0$, we obtain
\begin{align}
    \frac{1}{r^2}\left(r^2\tdPhi'\right)'-\frac{\ell(\ell+1)}{r^2}\tdPhi&=0,
    \\
    \frac{1}{r^2}\left(r^2\dpi'\right)'-\frac{\ell(\ell+1)}{r^2}\dpi&=0.
\end{align}
The solution that is regular at infinity is given by
\begin{align}
    \tdPhi=A_{\infty}r^{-\ell-1},
    \qquad 
    \dpi=B_{\infty}r^{-\ell-1},
    \label{bc:infty}
\end{align}
where $A_{\infty}$ and $B_{\infty}$ are constants.

Next, let us derive the solution in the vicinity of the center, where $x_0$ and $\tilde y_0$ are given respectively by eqs.~\eqref{soln-x0} and~\eqref{soln-y0}.
Taking the limit $\tilde y_0\sim x_0^2\gg 1$, we are allowed to ignore the terms with the coefficients $\xi$, $\hat\eta_0$, and $\hat\eta_3$ in eqs.~\eqref{pert-eq-1} and~\eqref{pert-eq-2}.
Substituting then $\mu\simeq \rho_c r^3/3$ and performing some manipulations, we obtain the decoupled equations valid in the vicinity of the center,
\begin{align}
    \frac{1}{r^2}\left(r^2\tdPhi'\right)'-\frac{\ell(\ell+1)}{r^2}\tdPhi&=S_\Phi,
    \\
    \frac{1}{r^2}\left(r^2\dpi'\right)'-\frac{\ell(\ell+1)}{r^2}\dpi&=S_\pi,
\end{align}
where
\begin{align}
    S_\Phi&=
    \frac{(1+3\alpha_H+3\beta_1)\delta\rho_{\ell m}+(\alpha_H+\beta_1)r\delta\rho'_{\ell m}}%
    {2(1+\alpha_H+\beta_1)}
    ,
    \\ 
    S_\pi&=\frac{\sigma}{r_c\rho_c^{1/2}}\cdot \frac{\sqrt{3}}{2}\left[
        \sqrt{c_0}\delta\rho_{\ell m}
        +\frac{(\alpha_H+\beta_1)(1+\alpha_H+\beta_1)}{2(\alpha_H+2\beta_1)}
        \frac{1}{\sqrt{c_0}}r\delta\rho'_{\ell m}
    \right],
\end{align}
with $c_0$ being the nonnegative number defined in eq.~\eqref{def:c0}.
Let us assume that near the center $\delta\rho_{\ell m}$ is of the form
\begin{align}
    \delta\rho_{\ell m}(r)\simeq d_{\ell m}r^n,
\end{align}
where $d_{\ell m}$ is a constant and $n$ is some positive number.
Then, we have $S_\Phi=s_\Phi d_{\ell m} r^n$ and $S_\pi=\left(\sigma/r_c\rho_c^{1/2}\right)s_\pi d_{\ell m}r^n$, where
\begin{align}
    s_\Phi&=\frac{(1+3\alpha_H+3\beta_1)+n(\alpha_H+\beta_1)}%
    {2(1+\alpha_H+\beta_1)},
    \\ 
    s_\pi&=\frac{\sqrt{3}}{2}\left[
        \sqrt{c_0}
        +\frac{(\alpha_H+\beta_1)(1+\alpha_H+\beta_1)}{2(\alpha_H+2\beta_1)}
        \frac{n}{\sqrt{c_0}}
    \right].
\end{align}
The particular solution is given by
\begin{align}
    \tdPhi^{\textrm{p}}&=-\frac{s_\Phi d_{\ell m}}{(\ell-n-2)(\ell+n+3)}r^{n+2},
    \\ 
    \dpi^{\textrm{p}}&=-\frac{\sigma}{r_c\rho_c^{1/2}}\cdot
    \frac{s_\pi d_{\ell m}}{(\ell-n-2)(\ell+n+3)}r^{n+2}.
\end{align}
This expression is valid for $\ell\neq n+2$.
When $\ell=n+2$, the particular solution is instead given by
\begin{align}
    \tdPhi^{\textrm{p}}&=\frac{s_\Phi d_{\ell m}}{(2n+5)}r^{n+2}\ln r,
    \\ 
    \dpi^{\textrm{p}}&=
    \frac{\sigma}{r_c\rho_c^{1/2}}\cdot\frac{s_\pi d_{\ell m}}{(2n+5)}r^{n+2}\ln r.
\end{align}
Adding the regular homogeneous solution, the general solution near the center is given by
\begin{align}
    \tdPhi=\tdPhi^{\textrm{p}}+A_0 r^{\ell},
    \qquad 
    \dpi=\dpi^{\textrm{p}}+B_0 r^{\ell},
    \label{bc:center}
\end{align}
where $A_0$ and $B_0$ are constants.

Equations~\eqref{pert-eq-1} and~\eqref{pert-eq-2} can be solved by the use of a standard shooting method.
We integrate them from the center outwards, subject to eq.~\eqref{bc:center},
and from some large $r$ inwards, subject to eq.~\eqref{bc:infty}.
We then match $\tdPhi$, $\dpi$, and their first derivatives at some intermediate $r$ to determine the four constants $A_\infty$, $B_\infty$, $A_0$, and $B_0$.

Before showing our numerical results, it is instructive to inspect four independent solutions to eqs.~\eqref{pert-eq-1} and~\eqref{pert-eq-2} in the region $R_0\lesssim r\ll r_{\textrm{V}}$, where $\delta\rho_{\ell m}$ can be ignored and $\mu\simeq \mu_0=\;$const.
In this regime, $\tdPhi$ and $\dpi$ are not decoupled:
\begin{align}
    &\left[
        \frac{1}{r^2}\left(r^2\tdPhi'\right)'-\frac{\ell(\ell+1)}{r^2}\tdPhi
    \right] 
    \notag \\ &
    =
    \frac{\alpha_H+2\beta_1}{(1+\alpha_H+\beta_1)^2}\left[
        \frac{2}{r^2}\left(r^2x_0\dpi'\right)'
    -\frac{\ell(\ell+1)}{r^3}\left(r^2x_0\right)'\dpi
    \right], 
    \\
    &2(1+\alpha_H+\beta_1)\biggl[
        \frac{2}{r^2}\left(r^2x_0\tdPhi'\right)'-\ell(\ell+1)
    \frac{(r^2x_0)'}{r^3}\tdPhi
    \notag \\ &
    +\frac{2}{r^2}\left(r^2\tilde y_0\dpi'\right)'-\ell(\ell+1)
    \frac{(r^2\tilde y_0)'}{r^3}\dpi
    \biggr]
    \notag \\ &
    -3\left[
        \frac{2}{r^2}\left(
        r^2x_0^2\dpi'
    \right)'-\frac{\ell(\ell+1)}{r^3}\left(r^2x_0^2\right)'\dpi
    \right]=0.
\end{align}
Assuming the solution of the form $\tdPhi\propto r^{\gamma}$ and $\dpi\propto r^{\gamma+3/2}$, we find four independent solutions of the form
\begin{align}
    \left(\tdPhi,\dpi\right)\propto
    \left(A_* r^\gamma,\frac{\sigma}{r_c\mu_0^{1/2}}r^{\gamma+3/2}\right),
\end{align}
where the powers are given by 
\begin{align}
    &\gamma_1=-\frac{1}{2}+\frac{\sqrt{D_+}}{2},
    \qquad 
    \gamma_2=-\frac{1}{2}-\frac{\sqrt{D_+}}{2},
    \notag \\ &
    \gamma_3=-\frac{1}{2}+\frac{\sqrt{D_-}}{2},
    \qquad
    \gamma_4=-\frac{1}{2}-\frac{\sqrt{D_-}}{2},
\end{align}
with
\begin{align}
    D_\pm=\ell(\ell+1)+\frac{5}{2}\pm\frac{3}{2}
    \sqrt{1-4\left(\frac{1+3\alpha_H+5\beta_1}{1-\alpha_H-3\beta_1}\right)\ell(\ell+1)
    +4\left(\frac{1+\alpha_H+\beta_1}{1-\alpha_H-3\beta_1}\right)[\ell(\ell+1)]^2},
\end{align}
and the relative amplitude $A_*$ is fixed as
\begin{align}
    A_*=\frac{3-2\ell(\ell+1)-4\gamma(\gamma+1)}%
    {(1+\alpha_H+\beta_1)(\ell+2\gamma)(\ell+1-2\gamma)}.
\end{align}
The relation~\eqref{rel:tildephi-phi} implies that for each of the solutions we have $\delta\Phi_{\ell m}\propto r^{\gamma}$.
For $|\alpha_H|,|\beta_1|\ll 1$, we obtain
\begin{align}
    &\gamma_1=\ell+\mathcal{O}(\alpha_H,\beta_1),
    \qquad 
    \gamma_2=-\ell-1+\mathcal{O}(\alpha_H,\beta_1),
    \notag \\ & 
    \gamma_3=-\frac{1}{2}+i\sqrt{\frac{(\ell+2)(\ell-1)}{2}}
    +\mathcal{O}(\alpha_H,\beta_1),
    \qquad 
    \gamma_4=-\frac{1}{2}-i\sqrt{\frac{(\ell+2)(\ell-1)}{2}}
    +\mathcal{O}(\alpha_H,\beta_1).
    \label{eq:powers-gamma}
\end{align}
Therefore, two of the four independent solutions have powers close to, but not equal to, the standard one ($\ell$ and $-\ell-1$) for small but nonvanishing $\alpha_H$ and $\beta_1$.
The other two are oscillatory.
Thus, there is no possibility for the solution in this regime to show exactly the expected $\sim r^{-\ell-1}$ behavior.
The solution is expressed as a linear combination of these four, and hence one would even expect a large deviation from the standard result due to the contamination of the latter two oscillatory components.

A similar nonstandard behavior of the gravitational potential in the nonlinear regime is seen without going beyond the Horndeski theory if one considers scalar-tensor theories with nonluminal gravitational wave propagation.
We provide a discussion on this point in appendix~\ref{app:Horn}.

\subsection{Numerical results}\label{subsec:example}

\begin{figure}
  \begin{minipage}{0.48\columnwidth}
    \centering
    \includegraphics[width=\linewidth]{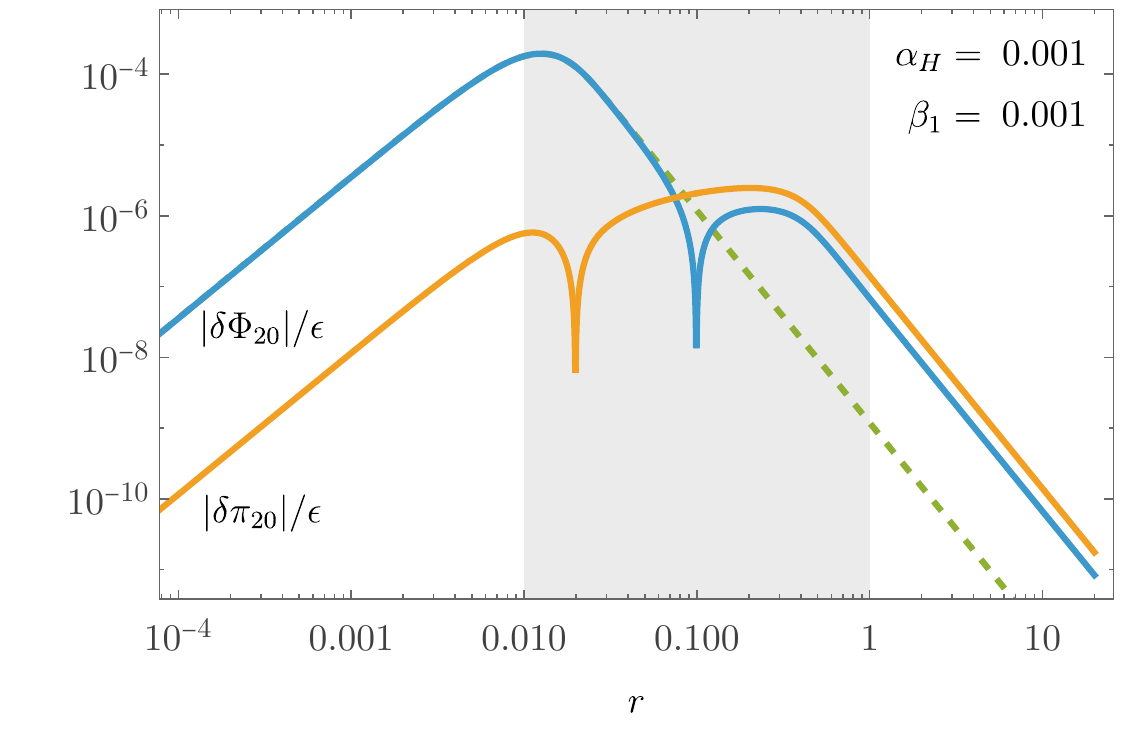}
  \end{minipage}
  \hfill
  \begin{minipage}{0.48\columnwidth}
    \centering
    \includegraphics[width=\linewidth]{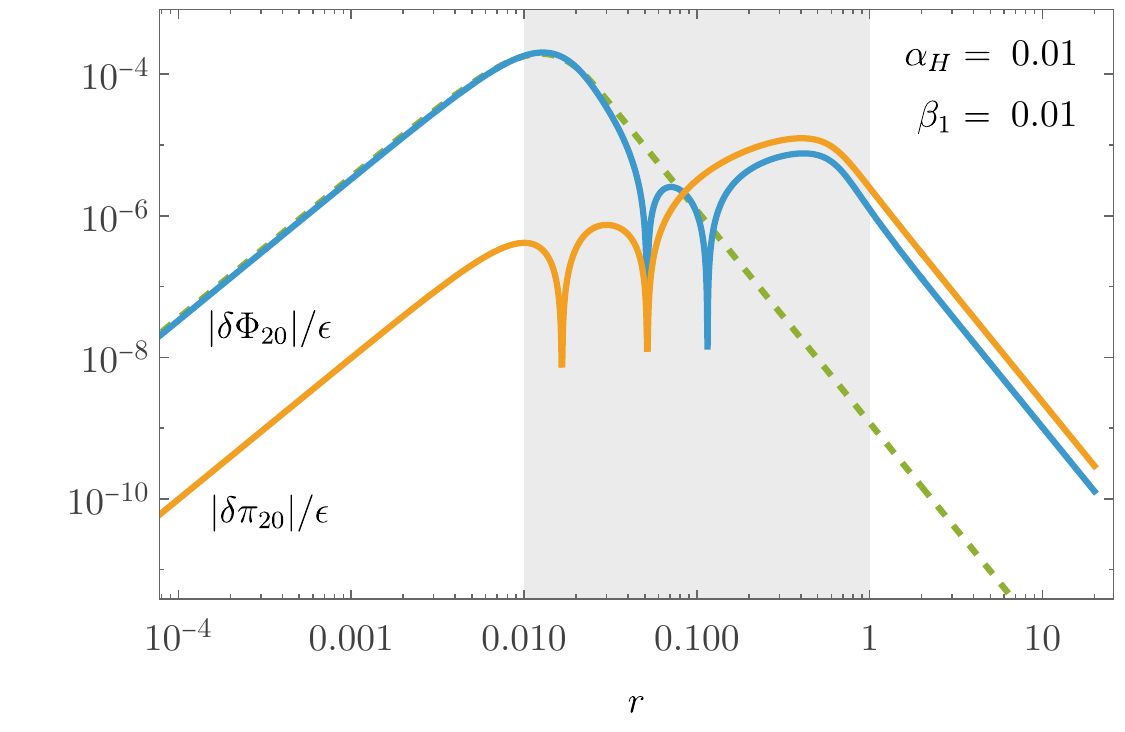}
  \end{minipage}
  \caption{Quadrupole moments $\delta\Phi_{20}$ (blue solid lines) and $\delta\pi_{20}$ (orange solid lines) are compared with the standard Newtonian result $\delta\Phi_{\textrm{N}20}$ (green dashed lines).
  The EFT parameters are given by $\hat\eta_0=1$, $\xi=1$, and $\hat\eta_3=0$ for both plots, while $\alpha_H$ and $\beta_1$ are chosen as shown in each plot.
  The parameters characterizing the density are chosen as $R_0=10^{-2}r_{\textrm{V}}$ and $\mu_0=10^{-3}R_0$, with $r_{\textrm{V}}=1$.
  The shaded region corresponds to $R_0\le r\le r_{\textrm{V}}$.}
  \label{fig:L=2}
\end{figure}

\begin{figure}
  \begin{minipage}{0.48\columnwidth}
    \centering
    \includegraphics[width=\linewidth]{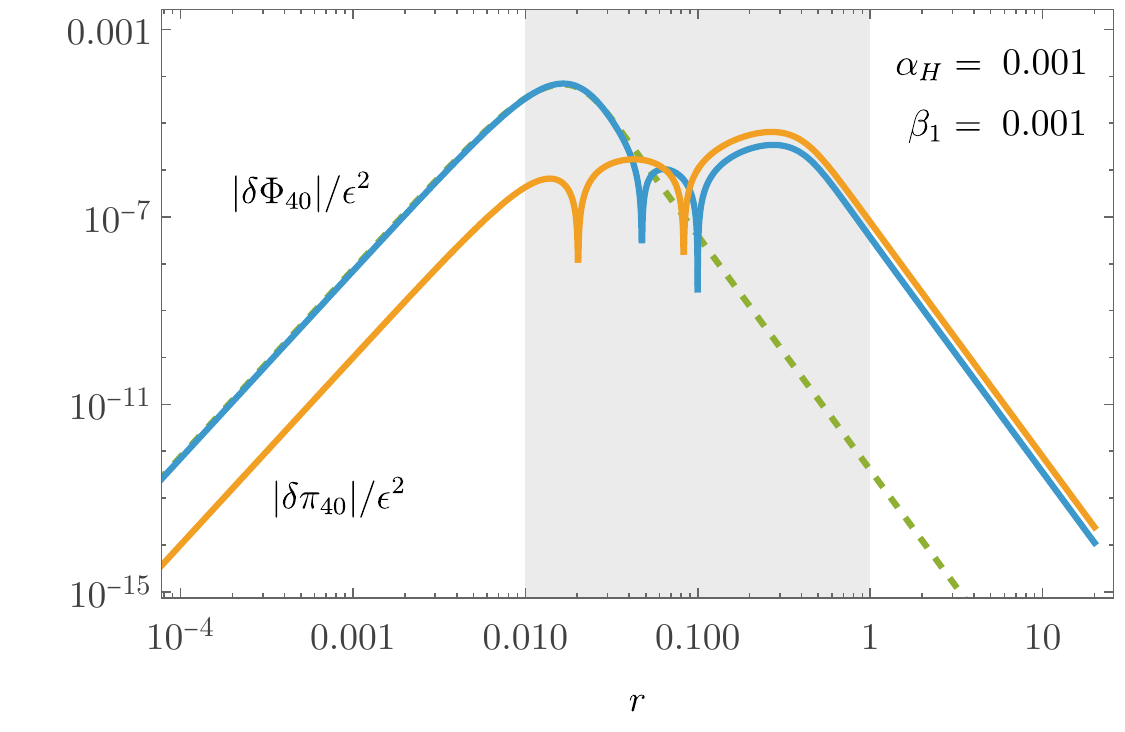}
  \end{minipage}
  \hfill
  \begin{minipage}{0.48\columnwidth}
    \centering
    \includegraphics[width=\linewidth]{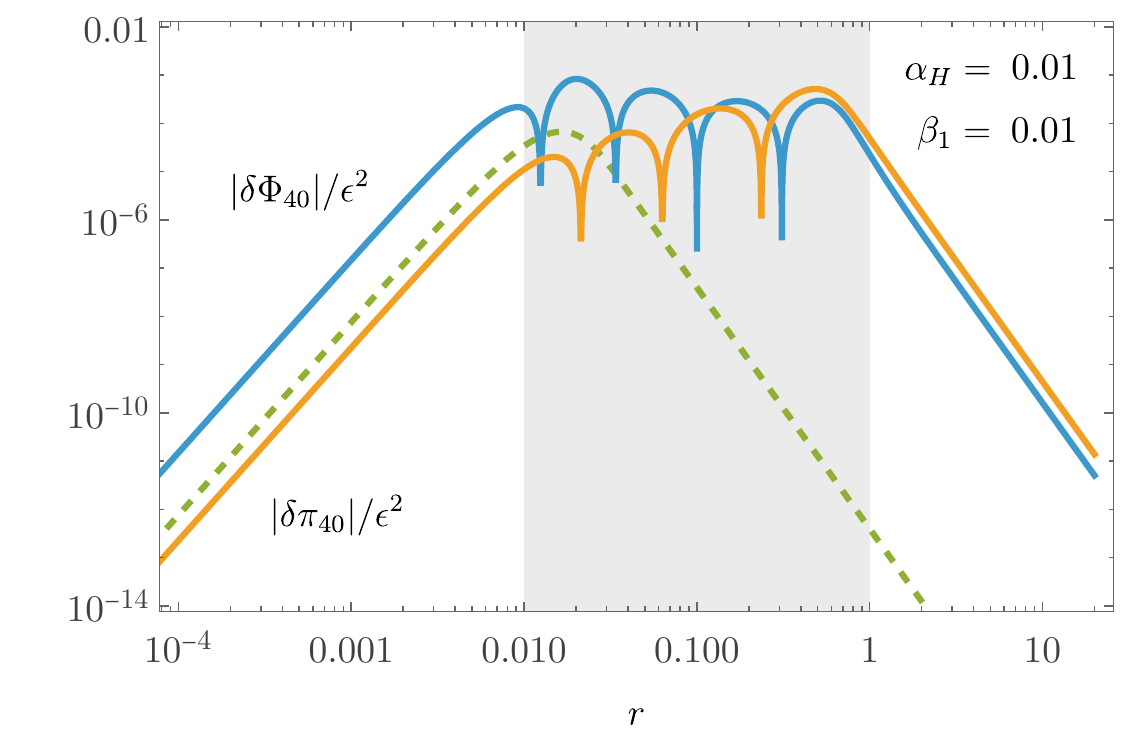}
  \end{minipage}
  \caption{Hexadecapole moments $\delta\Phi_{40}$ (blue solid lines) and $\delta\pi_{40}$ (orange solid lines) are compared with the standard Newtonian result $\delta\Phi_{\textrm{N}40}$ (green dashed lines).
  The EFT parameters are given by $\hat\eta_0=1$, $\xi=1$, and $\hat\eta_3=0$ for both plots, while $\alpha_H$ and $\beta_1$ are chosen as shown in each plot.
  The parameters characterizing the density are chosen as $R_0=10^{-2}r_{\textrm{V}}$ and $\mu_0=10^{-3}R_0$, with $r_{\textrm{V}}=1$.
  The shaded region corresponds to $R_0\le r\le r_{\textrm{V}}$.}
  \label{fig:L=4}
\end{figure}

As an example, let us consider a deformed density distribution of the form
\begin{align}
    \rho=\frac{4\mu_0}{\sqrt{\pi}R_0^3}
    \exp\left[-(x^2+y^2)/(b^{-1} R_0^2)-z^2/(b^2R_0^2)\right],
\end{align}
where $x$, $y$, and $z$ are the Cartesian coordinates and $b$ parametrizes the deviations from spherical symmetry, with $b=1$ corresponding to a spherically symmetric configuration.
If the deviations are small, one may write $b=1+\epsilon$ and expand $\rho$ in powers of $\epsilon$.
We find that
\begin{align}
    \rho= \rho_0(r)+
    \delta\rho_{20}(r)Y_{20}+\delta\rho_{40}(r)Y_{40}+\mathcal{O}(\epsilon^3),
\end{align}
where
\begin{align}
    \delta\rho_{20}(r)=
    \epsilon\frac{16\mu_0}{\sqrt{5}R_0^3}\cdot
    \frac{r^2}{R_0^2}e^{-r^2/R_0^2}+\mathcal{O}(\epsilon^2),
    \\ 
    \delta\rho_{40}(r)=
    \epsilon^2\frac{96\mu_0}{35R_0^3}\cdot
    \frac{r^4}{R_0^4}e^{-r^2/R_0^2}+\mathcal{O}(\epsilon^3).
\end{align}

For this density profile, we present in figures~\ref{fig:L=2} and~\ref{fig:L=4} the multipole moments of the gravitational potential $\delta\Phi_{20}$ and $\delta\Phi_{40}$ for two different sets of the EFT parameters.
We find that, for $R_0\lesssim r\lesssim 0.1\times r_{\textrm{V}}$ (where the Vainshtein mechanism works well for spherically symmetric configurations), the nonspherical parts of the gravitational potential behave differently from the corresponding standard Newtonian result~\eqref{eq:newton-result}.
Before evaluating $\delta\Psi_{\ell m}$, we can conclude that the multipole moments of the fifth force are not screened effectively.
The oscillatory behavior seen in the numerical results is due to the solutions with the imaginary powers $\gamma_3$ and $\gamma_4$.

\section{The special case: \texorpdfstring{$\alpha_H+2\beta_1=0$}{alphaH+2beta=0}}\label{sec:special-case}

\begin{figure}
  \begin{minipage}{0.48\columnwidth}
    \centering
    \includegraphics[width=\linewidth]{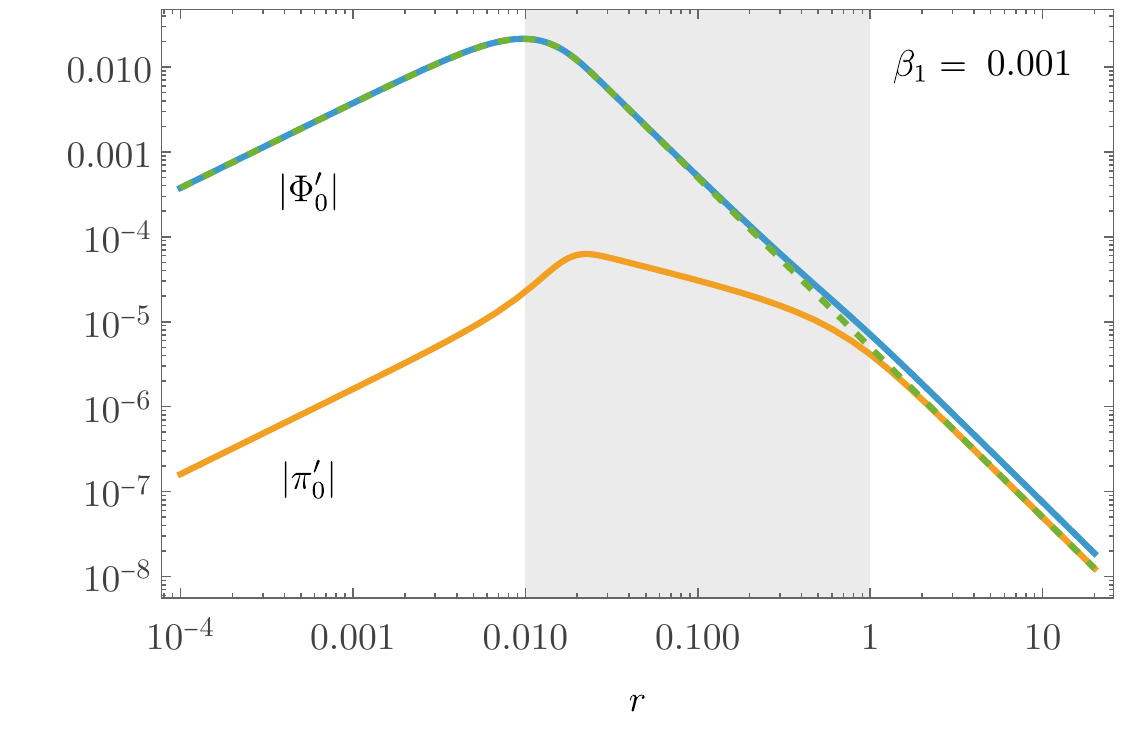}
  \end{minipage}
  \hfill
  \begin{minipage}{0.48\columnwidth}
    \centering
    \includegraphics[width=\linewidth]{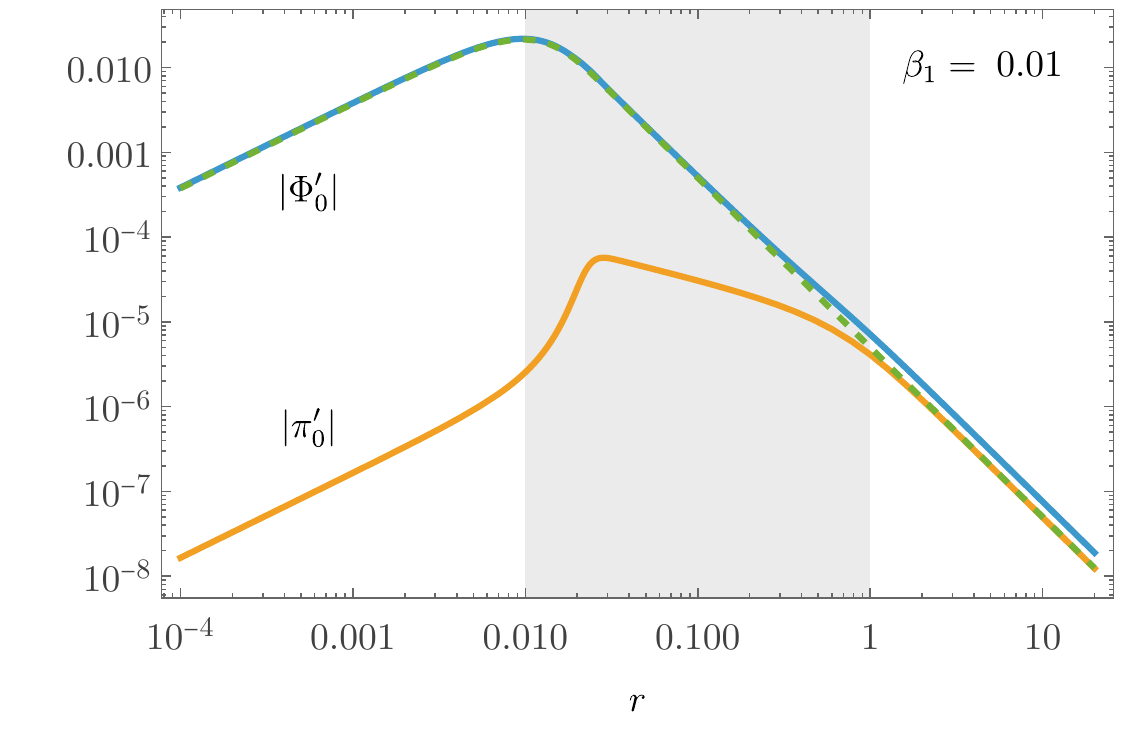}
  \end{minipage}
  \caption{The radial gradient of the gravitational potential (blue solid lines) is compared with the standard Newtonian result, $\mu(r)/2r^2$ (green dashed lines), in the special case of $\alpha_H+2\beta_1=0$.
  The radial gradient of the scalar field is also shown (orange solid lines).
  The EFT parameters are given by $\hat\eta_0=1$, $\xi=1$, and $\hat\eta_3=\xi(1-\beta_1)/2$ for both plots, while $\beta_1$ is chosen as shown in each plot.
  The parameters characterizing the density are chosen as $R_0=10^{-2}r_{\textrm{V}}$ and $\mu_0=10^{-3}R_0$ with $r_{\textrm{V}}=1$.
  The shaded region corresponds to $R_0\le r\le r_{\textrm{V}}$.}
  \label{fig:s-L=0}
\end{figure}

So far we have assumed that $\alpha_H+2\beta_1\neq 0$.
This is a crucial assumption, because the combination $\alpha_H+2\beta_1$ appears in the denominator in eqs.~\eqref{eq:spherical-Phi} and~\eqref{eq:spherical-Psi}.
Let us now turn to the discussion on the special case of $\alpha_H+2\beta_1=0$.
In this case, eqs.~\eqref{eq:y0} and~\eqref{eq:x0-y0} reduce respectively to
\begin{align}
    \tilde y_0=\frac{\xi}{2}x_0
    +\frac{\mathcal{A}}{2(1-\beta_1)}
    ,
    \label{eq:y0s}
\end{align}
and
\begin{align}
    &\hat\eta_0x_0-\frac{\xi}{2}(1-\beta_1)
    \mathcal{A}+\hat\eta_3x_0^2
    +\beta_1
    \frac{\left(r^3\mathcal{A}\right)'}{r^2}x_0=0.
    \label{eq:x0-y0s}
\end{align}
In what follows we assume that $\hat\eta_0>0$, $0<\beta_1<1$, and $\xi\hat\eta_3>0$.
The phenomenologically viable branch is then given by~\cite{Hirano:2019scf, Crisostomi:2019yfo}
\begin{align}
    x_0=\frac{1}{2\hat\eta_3}\left\{
        -\left[\hat\eta_0+\beta_1\frac{(r^3\mathcal{A})'}{r^2}\right]
        +\sqrt{\left[\hat\eta_0+\beta_1\frac{(r^3\mathcal{A})'}{r^2}\right]^2
        +2\xi\hat\eta_3(1-\beta_1)\mathcal{A}}
    \right\},
\end{align}
i.e.,
\begin{align}
    \pi_0'=\frac{r}{2r_c^2\hat\eta_3}\left[
        -\left(\hat\eta_0+r_c^2\beta_1\frac{\mu'}{r^2}\right)
        +\sqrt{\left(\hat\eta_0+r_c^2\beta_1\frac{\mu'}{r^2}\right)^2
        +2\xi\hat\eta_3(1-\beta_1)\frac{r_c^2\mu}{r^3}}
    \right].
\end{align}
In the linear regime ($\mathcal{A}\ll 1$), we have $x_0\simeq \xi(1-\beta_1)\mathcal{A}/2$.
In the opposite limit, $\mathcal{A}\gg 1$, the behavior of $x_0$ depends on whether $\mu'\neq 0$ or $\mu'=0$.
Inside material bodies, we have $\mu'\neq 0$, and hence 
\begin{align}
    x_0\simeq \frac{\xi(1-\beta_1)}{2\beta_1}\frac{\mu}{r\mu'}=\mathcal{O}(1),
\end{align}
yielding
\begin{align}
    \Phi_0'\simeq \frac{1}{2(1-\beta_1)}\frac{\mu}{r^2},
    \qquad 
    \Psi_0'\simeq (1-2\beta_1)\Phi_0'.
\end{align}
In the exterior region with $\mu=\mu_0=$ const but still in the nonlinear regime, $r\ll r_{\textrm{V}}=(r_c^2\mu_0)^{1/3}$, we have
\begin{align}
    x_0\simeq \frac{1}{\hat\eta_3}
    \sqrt{\frac{\xi\hat\eta_3(1-\beta_1)}{2}}\left(\frac{r_\textrm{V}}{r}\right)^{3/2}\gg 1,
\end{align}
and hence
\begin{align}
    \Phi_0'\simeq \left[\frac{1}{2(1-\beta_1)}-\frac{\xi\beta_1}{4\hat\eta_3}\right]\frac{\mu}{r^2},
    \qquad 
    \Psi_0'\simeq \left[\frac{1-2\beta_1}{2(1-\beta_1)}
    +\frac{\xi\beta_1}{4\hat\eta_3}\right]\frac{\mu}{r^2}.
\end{align}
Thus, only in the case where
\begin{align}
    \hat\eta_3=\frac{\xi}{2}(1-\beta_1)
    \label{fine-tune-paras}
\end{align}
is satisfied,
the two potentials coincide:
\begin{align}
    \Phi_0'\simeq \Psi_0'\simeq \frac{\mu}{2r^2}.
\end{align}
Thus, in this special case the partial breaking of Vainshtein screening occurs in a different way from the generic case: the effective gravitational constant inside material bodies is different by a factor of $(1-\beta_1)^{-1}$ from that in the exterior region~\cite{Hirano:2019scf, Crisostomi:2019yfo}. 
Furthermore, even in the exterior region we need a fine-tuning among the parameters in order for the two potentials to coincide.
For the Gaussian density profile~\eqref{gauss-density-0}, spherically symmetric solutions with $\beta_1=0.001$ and $\beta_1=0.01$ are shown in figure~\ref{fig:s-L=0}.

Let us consider slight deviations from spherical symmetry as in the previous section.
Equations~\eqref{pert-eq-1} and~\eqref{pert-eq-2} now reduce to
\begin{align}
    &\left[
        \frac{1}{r^2}\left(r^2\tdPhi'\right)'-\frac{\ell(\ell+1)}{r^2}\tdPhi
    \right] -\frac{\xi}{2}
    \left[
        \frac{1}{r^2}\left(r^2\dpi'\right)'-\frac{\ell(\ell+1)}{r^2}\dpi
    \right]
    -\frac{\delta\rho_{\ell m} }{2(1-\beta_1)}
    =0, 
    \label{pert-eq-1s}
\end{align}
and
\begin{align}
    &\hat\eta_0\left[
        \frac{1}{r^2}\left(r^2\dpi'\right)'-\frac{\ell(\ell+1)}{r^2}\dpi
        \right]
        -\frac{\xi}{2}(1-\beta_1)\delta\rho_{\ell m}
        \notag \\ &
        +\hat \eta_3
        \left[\frac{2}{r^2}\left(r^2x_0\dpi'\right)'
            -\frac{\ell(\ell+1)}{r^3}\left(r^2x_0\right)'\dpi\right]
    \notag \\ & 
    +\beta_1
    \left[
        \frac{1}{r^2}\left(r^3x_0\delta\rho_{\ell m}\right)'
    +\frac{r_c^2}{r^2}\left(\mu'\dpi'\right)'-\frac{\ell(\ell+1)r_c^2\mu'}{r^4}\dpi
    \right]
    =0.
    \label{pert-eq-2s}
\end{align}
In the linear regime, we may ignore $x_0$, $\mu'$, and $\delta\rho_{\ell m}$ in the above equations.
The two independent regular solutions are then given by
\begin{align}
    \tdPhi=A_\infty r^{-\ell-1},\qquad 
    \dpi=B_\infty r^{-\ell-1}.
\end{align}
In the vicinity of the center, eqs.~\eqref{pert-eq-1s} and~\eqref{pert-eq-2s} are decoupled in the limit $r_c^2\rho_c\sim (r_{\textrm{V}}/R_0)^3\gg 1$ to give
\begin{align}
    \frac{1}{r^2}\left(r^2\tdPhi'\right)'-\frac{\ell(\ell+1)}{r^2}
    \tdPhi&=\frac{d_{\ell m}r^n}{2(1-\beta_1)},
    \\ 
    \frac{1}{r^2}\left(r^2\dpi'\right)'-\frac{\ell(\ell+1)}{r^2}
    \dpi&=
    -\frac{n\xi}{6r_c^2\rho_c}\left(\frac{1-\beta_1}{\beta_1}\right)
    d_{\ell m}r^n,
\end{align}
where we assumed $\delta\rho_{\ell m}=d_{\ell m}r^n$ there.
Thus, the regular solution near the center is of the form
\begin{align}
    \tdPhi&=-\frac{d_{\ell m}r^{n+2}}{2(\ell-n-2)(\ell+n+3)(1-\beta_1)}+A_0r^\ell,
    \\ 
    \dpi&=\frac{n(1-\beta_1)\xi d_{\ell m}r^{n+2}}{6(\ell-n-2)(\ell+n+3)\beta_1r_c^2\rho_c}
    +B_0r^\ell.
\end{align}
(For simpllicity, we omit the solution for $\ell=n+2$.)
We are now ready to solve eqs.~\eqref{pert-eq-1s} and~\eqref{pert-eq-2s} numerically using a shooting method.

\begin{figure}
  \begin{minipage}{0.48\columnwidth}
    \centering
    \includegraphics[width=\linewidth]{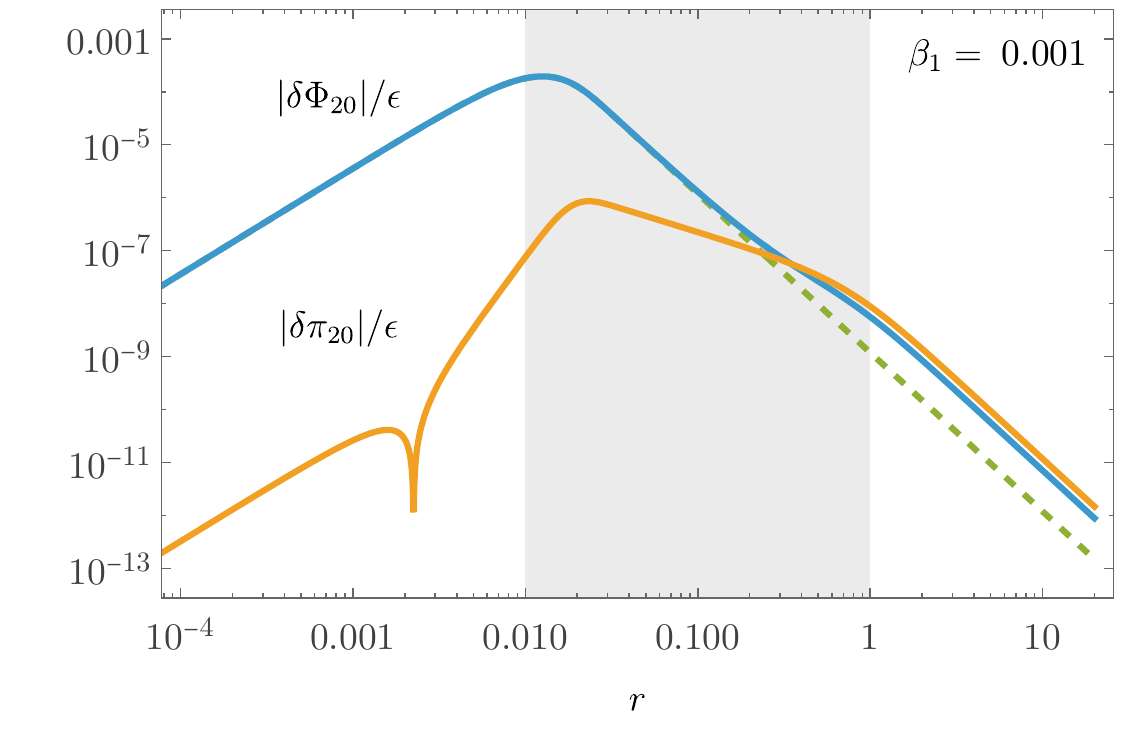}
  \end{minipage}
  \hfill
  \begin{minipage}{0.48\columnwidth}
    \centering
    \includegraphics[width=\linewidth]{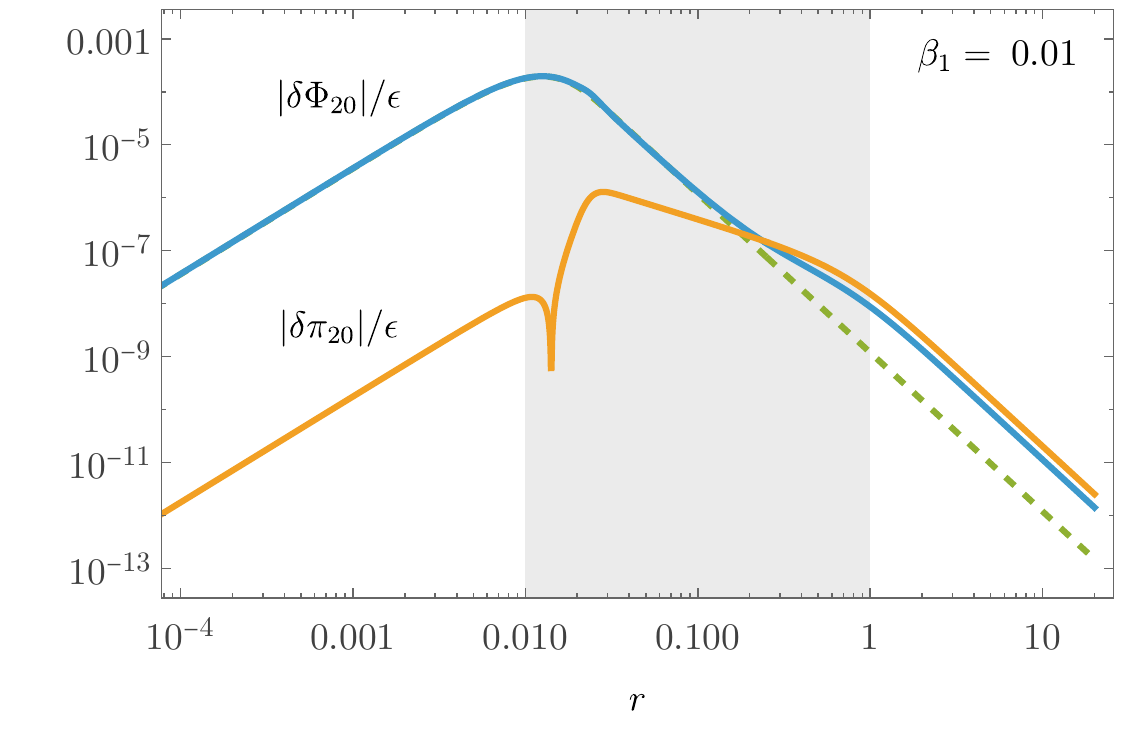}
  \end{minipage}
  \caption{Quadrupole moments $\delta\Phi_{20}$ (blue solid lines) and $\delta\pi_{20}$ (orange solid lines) are compared with the standard Newtonian result $\delta\Phi_{\textrm{N}20}$ (green dashed lines) in the case of $\alpha_H+2\beta_1=0$.
  The EFT parameters are given by $\hat\eta_0=1$, $\xi=1$, and $\hat\eta_3=\xi(1-\beta_1)/2$ for both plots, while $\beta_1$ is chosen as shown in each plot.
  The parameters characterizing the density are chosen as $R_0=10^{-2}r_{\textrm{V}}$ and $\mu_0=10^{-3}R_0$, with $r_{\textrm{V}}=1$.
  The shaded region corresponds to $R_0\le r\le r_{\textrm{V}}$.}
  \label{fig:s-L=2}
\end{figure}
\begin{figure}
  \begin{minipage}{0.48\columnwidth}
    \centering
    \includegraphics[width=\linewidth]{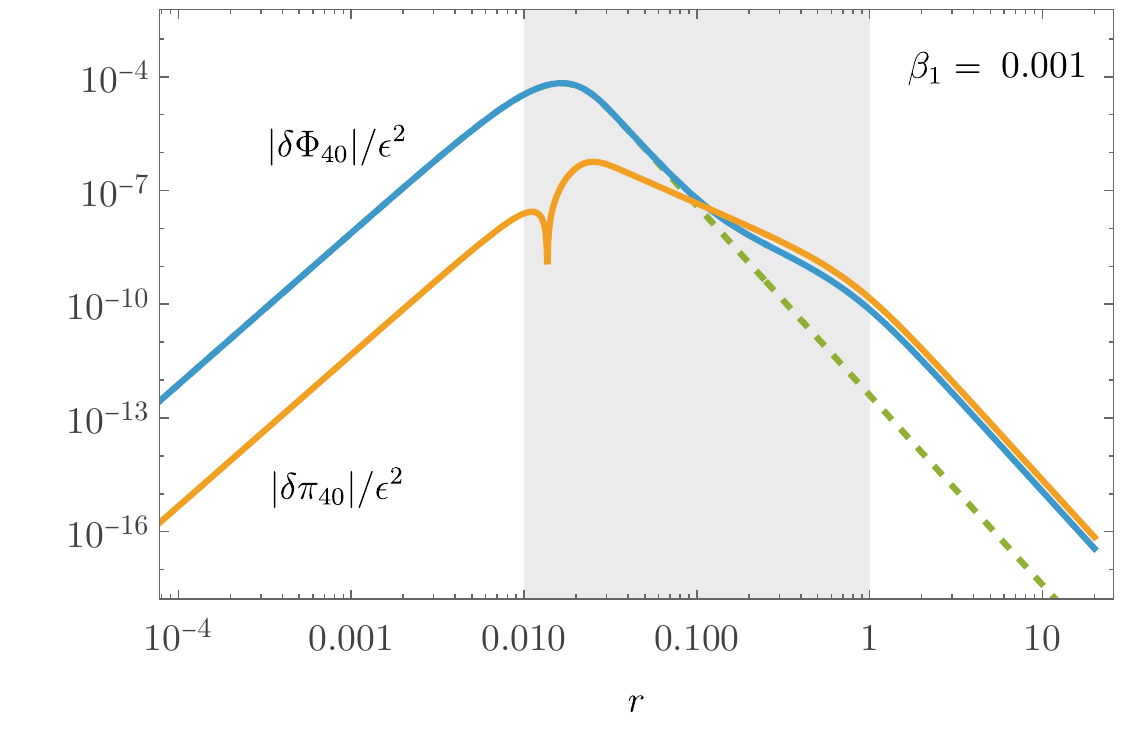}
  \end{minipage}
  \hfill
  \begin{minipage}{0.48\columnwidth}
    \centering
    \includegraphics[width=\linewidth]{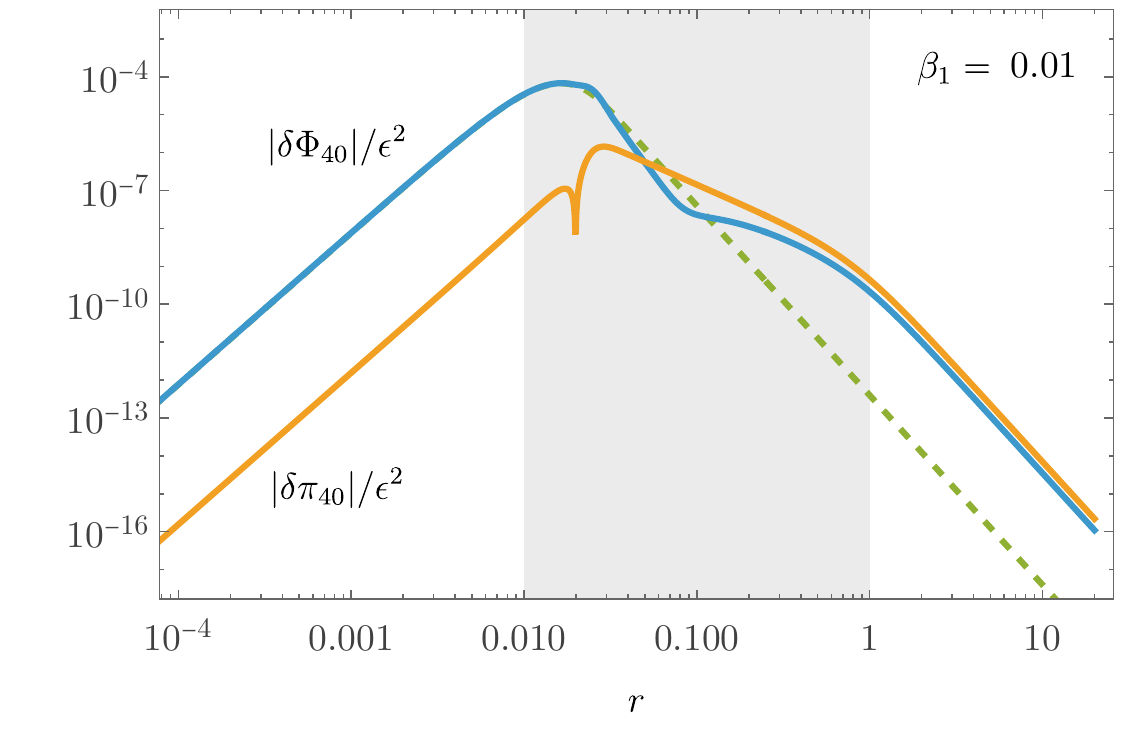}
  \end{minipage}
  \caption{Hexadecapole moments $\delta\Phi_{40}$ (blue solid lines) and $\delta\pi_{40}$ (orange solid lines) are compared with the standard Newtonian result $\delta\Phi_{\textrm{N}40}$ (green dashed lines) in the case of $\alpha_H+2\beta_1=0$.
  The EFT parameters are given by $\hat\eta_0=1$, $\xi=1$, and $\hat\eta_3=\xi(1-\beta_1)/2$ for both plots, while $\beta_1$ is chosen as shown in each plot.
  The parameters characterizing the density are chosen as $R_0=10^{-2}r_{\textrm{V}}$ and $\mu_0=10^{-3}R_0$, with $r_{\textrm{V}}=1$.
  The shaded region corresponds to $R_0\le r\le r_{\textrm{V}}$.}
  \label{fig:s-L=4}
\end{figure}

Our numerical results are displayed in figures~\ref{fig:s-L=2} and~\ref{fig:s-L=4}.
In contrast to the generic case discussed in the previous section, we do not see characteristic large deviations from the standard Newtonian results for $R_0\lesssim r\lesssim 0.1\times r_{\textrm{V}}$ when $\alpha_H+2\beta_1=0$, though $\delta\Phi_{\ell m}$ is seen to deviate away from $\delta\Phi_{\textrm{N}\ell m}$ to some extent for $\ell=4$ and $\beta_1=0.01$.
To better understand the numerical results, let us derive the analytic expressions for $\tdPhi$ and $\dpi$ valid in the range $R_0\lesssim r\ll r_{\textrm{V}}$.
In this region, eqs.~\eqref{pert-eq-1s} and~\eqref{pert-eq-2s} reduce respectively to
\begin{align}
    \frac{1}{r^2}\left(r^2\tdPhi'\right)'-\frac{\ell(\ell+1)}{r^2}\tdPhi
    &=\frac{3\xi}{8}\left[
        \frac{2}{r}\dpi'-\frac{\ell(\ell+1)}{r^2}\dpi
    \right],
    \\
    \dpi''+\frac{\dpi'}{2r}-\frac{\ell(\ell+1)}{4r^2}\dpi&=0,
\end{align}
and the general solution is given by
\begin{align}
    \tdPhi&=C_1r^\ell+C_2r^{-\ell-1}+\frac{\xi}{2}\dpi,
    \\
    \dpi&=C_3r^{(\ell+1)/2}+C_4r^{-\ell/2}.
\end{align}
The form of $\dpi$ is the same as that in the case of a cubic galileon in flat space (see appendix~\ref{app:flat-space-galileon}).
The gravitational potentials can be calculated from the relations
\begin{align}
    \delta\Phi_{\ell m}&=\tdPhi+\frac{\beta_1}{1-\beta_1}r_c^2\pi_0'\dpi'
    \notag \\ &
    =C_1r^\ell+C_2r^{-\ell-1}+\frac{\xi}{2}\dpi+\frac{\beta_1}{1-\beta_1}x_0r\dpi',
    \\
    \delta\Psi_{\ell m}&=(1-2\beta_1)\delta\Phi_{\ell m}
    +\frac{\eta_2}{4}\dpi-2\beta_1x_0r\dpi'.
    \label{s-case-psi-phi}
\end{align}
We see that $\delta\Phi_{\ell m}$ is composed of the components with the standard Newtonian behavior ($r^\ell$, $r^{-\ell-1}$) and the nonstandard ones that come from $\dpi$ and $\pi_0'\dpi'$.
If the latter nonstandard contributions are suppressed, we may expect the standard $\sim r^{-\ell-1}$ behavior.
This is indeed the case for our numerical solutions with $(\ell,\beta_1)=(2,0.001), (2,0.01)$, and $(4, 0.001)$.
However, it then follows from eq.~\eqref{s-case-psi-phi} that $\delta\Psi_{\ell m}=(1-2\beta_1)\delta\Phi_{\ell m}$, showing that the two potentials do not coincide and hence the screening mechanism fails to operate.

\section{Conclusions}\label{sec:disc}

In this paper, we have investigated whether the Vainshtein mechanism operates to screen the fifth force beyond spherical symmetry in degenerate higher-order scalar-tensor (DHOST) theories and the effective field theory (EFT) of dark energy, focusing on the role of the EFT parameters that characterize the ``beyond Horndeski'' part of the Lagrangian (denoted conventionally as $\alpha_H$ and $\beta_1$ in the $\alpha$-basis EFT~\cite{Langlois:2017mxy, Crisostomi:2017lbg}).
It has been shown, under the assumption of spherical symmetry, that the fifth force is suppressed and the standard Newtonian gravitational potential is reproduced inside the Vainshtein radius and in the exterior of a gravitational source~\cite{Kobayashi:2014ida, Dima:2017pwp, Langlois:2017dyl, Crisostomi:2017lbg}.
However, this result relies on a delicate interplay among multiple terms, and it is not obvious whether the same conclusion persists when the assumption of spherical symmetry is relaxed.
We have addressed this question by considering the case where the matter distribution deviates only slightly from spherical symmetry.

In the simpler setup of a flat-space galileon coupled directly to a source, one can show that $|\delta\Phi_{\textrm{N}\ell m}|\gg |\dpi|$, where $\delta\Phi_{\textrm{N}\ell m}$ is the multipole moment of the Newtonian potential and $\dpi$ is that of the galileon field, and thus the fifth force is screened even in the absence of spherical symmetry (see appendix~\ref{app:flat-space-galileon}).
In the EFT of dark energy, however, we have found that the field equations do not admit the solution $\delta\Phi_{\ell m}=\delta\Phi_{\textrm{N}\ell m}\propto r^{-\ell-1}$ in the exterior of the source for a generic choice of the ``beyond Horndeski'' parameters $\alpha_H$ and $\beta_1$.
Instead, we have revealed a characteristic oscillatory behavior of $\delta\Phi_{\ell m}$.
The Vainshtein mechanism thus fails to operate.
We have the special case satisfying $\alpha_H+2\beta_1=0$ in which the structure of the field equations governing Vainshtein screening is different.
Also, in this case, we have shown that the standard behavior of the gravitational potentials cannot be reproduced outside a source.
Thus, we conclude that the Vainshtein mechanism does not work for the multipole moments of the gravitational potentials in the exterior of a nonspherical source if there are beyond Horndeski terms in the Lagrangian.
Note that our findings apply to the influence of mass multipole moments on the gravitational potential, and hence are not in contradiction to the results of ref.~\cite{Anson:2020fum}, in which it was shown that the Vainshtein mechanism works for the frame-dragging effect of a slowly rotating body.

The failure of Vainshtein screening for the multipole moments would allow us to put constraints on the EFT parameters $\alpha_H$ and $\beta_1$ for instance from the measurements of the gravitational moments of the Sun and the Earth, which are quantified by the coefficients $J_\ell$ defined by the expression $\delta\Phi=-\Phi_0\sum J_\ell (R_*/r)^\ell P_\ell(\cos\theta)$.
Here, $R_*$ is the mean radius of the source body and $P_\ell$ is the Legendre polynomial of degree $\ell$.
For the Sun, indirect determination of $J_\ell$ has been made through the inference of internal rotation based on helioseismology~\cite{Pijpers:1998eb, Mecheri:2004eb, Mecheri:2021eb}, while they have been measured more directly through the precession of Mercury's perihelion~\cite{Park:2017zgd, Genova:2018mjp}, giving values consistent with each other.
For the Earth, $J_2$ (and its temporal variation) has been determined from satellite geodesy~\cite{yoder1983secular, Pavlis2021, Cheng2013}.
Our results imply that in the EFT of dark energy the coefficients $J_\ell$ effectively depend on $r$ and deviate from the standard values since $\delta\Phi_{\ell m}$ no longer obeys the $\sim r^{-\ell-1}$ law and its relation to the internal structure of the source is complicated due to the mixing with the scalar field.
It would be interesting to confront the EFT of dark energy with these experiments.
To do so, we need precise computations of the gravitational fields of the Sun and the Earth based on a more realistic astrophysical setup in the EFT of dark energy.

\acknowledgments
The work of TK was supported by JSPS KAKENHI Grant No.~JP25K07308.
The work of TT was supported by the Sasakawa Scientific Research Grant from The Japan Science Society.

\appendix

\section{The effective theory from the class-Ia DHOST action}\label{app:eft-derivation}

For completeness, we review the connection between the effective theory~\eqref{effLag} and the action for the class-Ia quadratic DHOST theories.
The action we consider is given by~\cite{Langlois:2015cwa}
\begin{align}
        S&=\int\D^4x\sqrt{-g}\left[P+Q\Box\phi+f
        \mathcal{R}+\sum_{I=1}^5A_IL_I\right],\label{action:dhost}
\end{align}
where, using the notations $\phi_\mu=\nabla_\mu\phi$ and $\phi_{\mu\nu}=\nabla_\mu\nabla_\nu\phi$,
\begin{align}
        L_1&:=\phi_{\mu\nu}\phi^{\mu\nu},\qquad 
        L_2:=(\Box\phi)^2,\qquad L_3:=\Box\phi\phi^\mu\phi_{\mu\nu}\phi^\nu,
        \notag \\ 
        L_4&:=\phi^\mu\phi_{\mu\nu}\phi^{\nu\lambda}\phi_\lambda,
        \qquad L_5:=\left(\phi^\mu\phi_{\mu\nu}\phi^\nu\right)^2,\label{lag:sup1}
\end{align}
$\mathcal{R}$ is the Ricci scalar,
and $P$, $Q$, $f$, and $A_I$ are functions of the scalar field $\phi$ and $X:=-g^{\mu\nu}\phi_\mu\phi_\nu/2$.
We focus on a subset of theories satisfying $A_1+A_2=0$, because it constitutes a physically interesting class of theories (class-Ia DHOST theories)~\cite{deRham:2016wji, Langlois:2017mxy}.

We consider fluctuations around a cosmological background, and write the scalar field as
\begin{align}
    \phi=\bar{\phi}(t)+\delta\phi(t,\Vec{x}),\label{ap:scalar-pert}
\end{align}
and the metric in the Newtonian gauge as
\begin{align}
    \D s^2=-[1+2\Phi(t,\Vec{x})]\D t^2+a^2(t)[1-2\Psi(t,\Vec{x})]\delta_{ij}\D x^i\D x^j.
    \label{ap:metric-pert-N}
\end{align}
In what follows, we denote just by $\phi$ the homogeneous part of the scalar field, omitting the bar.

Substituting the scalar field~\eqref{ap:scalar-pert} and the metric~\eqref{ap:metric-pert-N} into eq.~\eqref{action:dhost} and expanding the action under the quasi-static approximation, we obtain the effective action of the form~\cite{Dima:2017pwp}
\begin{align}
        S &=\int\D^4x\frac{M^2a}{2}\biggl[
                \left(
                        c_1\Phi+c_2\Psi+c_3\delta\phi
                \right)\nabla^2\delta\phi + c_4\Psi\nabla^2\Phi +c_5 \Psi\nabla^2\Psi 
                +c_6\Phi\nabla^2\Phi 
                \notag \\ & \quad 
                +\left(c_7\dot\Psi+c_8\dot\Phi+c_9\ddot{\delta\phi}\right)\nabla^2\delta\phi
                +\frac{b_1}{a^2}\mathcal{L}_3^{\textrm{Gal}}
                +\frac{1}{a^2}\left(b_2 \Phi+b_3\Psi \right)\mathcal{E}_3^{\textrm{Gal}}
                \notag \\ & \quad
                +\frac{1}{a^2}\left(b_4\nabla_i\Psi+b_5\nabla_i\Phi +b_6\nabla_i\dot{\delta\phi} \right) 
                \nabla_j\delta\phi\nabla_i\nabla_j\delta\phi
                \notag \\ & \quad 
                +\frac{1}{a^4}\left(
                        d_1\mathcal{L}_4^{\textrm{Gal}}
                        +d_2\nabla_i\delta\phi\nabla_j\delta\phi \nabla_i\nabla_k\delta\phi
                        \nabla_j\nabla_k\delta\phi 
                        \right)
        \biggr],
        \label{ap:efective-action}
\end{align}
where
\begin{align}  
        &\mathcal{L}_3^{\textrm{Gal}}:=-\frac{1}{2}(\nabla\delta\phi)^2\nabla^2\delta\phi,
        \qquad
        \mathcal{E}_3^{\textrm{Gal}}:=(\nabla^2\delta\phi)^2
        -\nabla_i\nabla_j\delta\phi\nabla_i\nabla_j\delta\phi,
        \notag \\ & 
        \mathcal{L}_4^{\textrm{Gal}}:=-\frac{1}{2}(\nabla\delta\phi)^2\mathcal{E}_3^{\textrm{Gal}}.
\end{align}
We introduce the following time-dependent functions:
\begin{align}
    &M^2:=2\left(f+2XA_1\right),
    \qquad 
    M^2H\alpha_M:= \frac{\D}{\D t}M^2,
    \notag \\ &
    M^2H\alpha_B:= \left(3 f_{X}+2 X f_{XX}+3A_1+2XA_{1X}+3XA_{3}+2X^2A_{3X}-2XA_{4}+4X^2A_5\right)\dot{\phi}\ddot{\phi} 
    \notag \\ &\qquad \qquad \quad
    -2H\left(f_{X}- A_{1}+2 XA_{1X}-6 X A_{1X}+3 XA_{3}\right)X
    \notag \\ &\qquad \qquad \quad 
    +\left(f_{\phi}+2Xf_{\phi X}+X Q_{X}\right)\dot{\phi},   
    \notag \\ & 
    M^2\alpha_T:=-4XA_1,
        \qquad 
        M^2\alpha_H:=-4X\left(f_X+A_1\right),
        \qquad 
        M^2\alpha_{V}:=4X\left[f_X+2\left(A_1+XA_{1X}\right)\right],
        \notag \\ 
        &M^2\beta_1:=2X\left(f_X+A_1+XA_3\right),
        \qquad 
        M^2\beta_3:=-8X\left(f_X+A_1-XA_4\right)
        ,
        \label{defs:eftparas}
\end{align}
where $H:=\dot a/a$ and a dot stands for differentiation with respect to $t$.
The functions of $\phi$ and $X$ in the right hand sides are evaluated at the cosmological background.
In terms of these functions, the coefficients in the action~\eqref{ap:efective-action} are given by
\begin{align}
        &c_1= -\frac{2}{\dot\phi}\left\{H\left[2\alpha_B-2\alpha_H+\beta_3(1+\alpha_M)\right]+\dot\beta_3\right\},
        \notag \\
        &c_2= \frac{4}{\dot\phi}
        \left\{H\left[\alpha_M+\alpha_H(1+\alpha_M)-\alpha_T\right]+\dot\alpha_H\right\},
        \notag \\ 
        &c_3= -\frac{1}{4X}\biggl\{
        H^2\left[4\alpha_B(1+\alpha_M)-4(\alpha_H+\alpha_M+\alpha_H\alpha_M-\alpha_T)+(1+\alpha_M)^2\beta_3\right]
        \notag \\ &\quad\quad
        +\dot{H}\left(4+4\alpha_B-4\alpha_H+\beta_3+\alpha_M\beta_3\right)
        +H\left[4\dot{\alpha}_B-4\dot{\alpha}_H+\beta_3\dot{\alpha}_M+2(1+\alpha_M)\dot{\beta}_3\right] 
        +\ddot{\beta}_3\biggr\}
        \notag \\ &\quad\quad 
        +\left[H(1+\alpha_M)(4\beta_1+\beta_3)+4\dot\beta_1+\dot\beta_3\right] 
        \frac{\ddot\phi}{2\dot\phi X} 
        -(4\beta_1+\beta_3)\frac{\ddot\phi^2}{4X^2}
        -\frac{\bar\rho+\bar p}{2M^2X},
        \notag \\
        &c_4=4(1+\alpha_H),
        \qquad 
        c_5=-2(1+\alpha_T),
        \qquad 
        c_6=-\beta_3,
        \notag \\ &
        c_7=\frac{4\alpha_H}{\dot\phi},
        \qquad 
        c_8=-\frac{2}{\dot\phi}(2\beta_1+\beta_3),
        \qquad 
        c_9=\frac{1}{2X}(4\beta_1+\beta_3),
        \notag \\ &
        b_1=\frac{1}{6X\dot{\phi}}
        \biggl\{H \left[12\alpha_{B}-3 \alpha_{H} (\alpha_{M}+3)+3 \alpha_{M} (\alpha_{V}
        -8 \beta_{1}-2)+9 \alpha_{T}-3 \alpha_{V}\right]
        \notag \\ &\quad\quad        
        -3\dot{\alpha}_{H}+3 \left(\dot{\alpha}_{V}-8 \dot{\beta}_{1}\right)\biggr\}
        +3 (4 \beta_{1}+\beta_{3})\frac{\ddot{\phi}}{6X^2},
        \notag \\ &
        b_2=
        \frac{1}{2X}\left(-\alpha_H-4\beta_1+\alpha_{V}\right),
        \qquad 
        b_3=\frac{\alpha_T}{2X},
        \qquad
        b_4=-\frac{2\alpha_H}{X},
        \qquad 
        b_5=\frac{1}{X}\left(2\beta_1+\beta_3\right),
        \notag \\ &
        b_6=-\frac{1}{\dot\phi X}\left(4\beta_1+\beta_3\right), 
        \qquad
        d_1=-\frac{1}{2X}(b_2+b_3),
        \qquad 
        d_2=\frac{1}{4X^2}\left(4\beta_1+\beta_3\right),
\end{align}
with $\bar\rho$ and $\bar p$ being respectively the energy density and pressure of the cosmological background.

In the main text, we use the dimensionless scalar-field fluctuation $\pi$ defined as
\begin{align}
    \pi:=\frac{\delta\phi}{r_c\dot\phi},
\end{align}
where $r_c$ is a parameter of the order of the cosmological horizon scale ($\sim H^{-1}$).
We neglect the time variation of the scale factor and set $a=1$, while keeping $r_c\sim H^{-1}$ finite.
We thus find the explicit connection between the effective field theory of dark energy and the action for DHOST theories.
Requiring that the speed of gravitational waves is equal to that of light amounts to imposing $\alpha_T=0$ and $\alpha_H=-\alpha_V$.

\section{Flat-space galileon}\label{app:flat-space-galileon}

To highlight the main results of the present paper derived from the EFT of dark energy, in this appendix we compute the multipole moments of a flat-space galileon coupled to the energy density of nonrelativistic matter.
The Lagrangian for a flat-space galileon is given by
\begin{align}
    \mathcal{L}=\frac{M^2}{2}
    \left(\eta_0\pi\nabla^2\pi +r_c^2\eta_3\mathcal{L}_3^{\textrm{Gal}}
    +r_c^4\alpha\mathcal{L}_4^{\textrm{Gal}}\right)
    -\pi \rho,
\end{align}
where $\eta_0$, $\eta_3$, and $\alpha$ are positive dimensionless constants.
The field equation is obtained as
\begin{align}
    \eta_0\nabla^2\pi+\frac{r_c^2\eta_3}{2}\mathcal{E}_3^{\textrm{Gal}}
    +\frac{r_c^4\alpha}{2}\mathcal{E}_4^{\textrm{Gal}}
    =\rho,
\end{align}
where, as in the main text, we redefine $\rho/M^2$ as $\rho$.

In terms of the dimensionless function $x_0(r):=r_c^2\pi_0'/r$, the spherically symmetric part of the field equation reads
\begin{align}
    \eta_0x_0+\eta_3x_0^2+\alpha x^3_0=\mathcal{A}.
\end{align}
In the Vainshtein regime, $\mathcal{A}\gg 1$, we have $x_0\gg 1$, yielding
\begin{align}
    \pi_0'\simeq 
    \begin{cases}
        \displaystyle{\frac{\mu_0}{\alpha^{1/3}r_{\textrm{V}}^2}\qquad 
        \qquad\qquad ~(\alpha\neq 0)}
        \\
        \displaystyle{\frac{1}{\eta_3^{1/2}}\frac{\mu_0}{r^2}\left(\frac{r}{r_{\textrm{V}}}\right)^{3/2}
        \qquad (\alpha=0)}
    \end{cases},
    \label{eq:flat-g-soln-Vianshtein}
\end{align}
where we have assumed that $\mu=\mu_0=$ const and hence $\mathcal{A}=r_c^2\mu_0/r^3$ in this regime.
This result should be compared with the spherically symmetric solution to the standard Poisson equation for the Newtonian potential, $\nabla^2\Phi=\rho/2$, i.e., $\Phi_0'=\mu_0/2r^2$.
We see that the galileon-mediated fifth force is suppressed, $\pi_0'\ll \Phi_0'$, for $r\ll r_{\textrm{V}}$.

A small deviation from spherical symmetry, $\dpi$, obeys
\begin{align}
    &\eta_0\left[\frac{1}{r^2}\left(r^2\dpi'\right)'-\frac{\ell(\ell+1)}{r^2}\dpi\right]
    +\eta_3\left[
        \frac{2}{r^2}\left(r^2x_0\dpi'\right)'-\frac{\ell(\ell+1)}{r^3}
        \left(r^2x_0\right)'\dpi
    \right]
    \notag \\ & 
    +\alpha\left[
        \frac{3}{r^2}\left(r^2x_0^2\dpi'\right)'
        -\frac{3\ell(\ell +1)}{2r^3}\left(r^2x_0^2\right)'\dpi
    \right]
    =\delta\rho_{\ell m}.\label{eq:flat-g-pert}
\end{align}
At infinity, $x_0\ll 1$ and $\delta\rho_{\ell m}=0$, and hence we impose
\begin{align}
    \dpi=A_\infty r^{-\ell-1}.
\end{align}
Suppose that the density is of the form $\delta\rho_{\ell m}=d_{\ell m}r^n$ near the center.
Since $x_0\gg 1$ in the vicinity of the center, we have
\begin{align}
    \frac{1}{r^2}\left(r^2\dpi'\right)'-\frac{\ell(\ell+1)}{r^2}\dpi 
    =
    \begin{cases}
        \displaystyle{\frac{d_{\ell m}r^n}{(3\alpha)^{1/3}(r_c^2\rho_c)^{2/3}}
        \qquad (\alpha\neq 0)}
        \\
        \displaystyle{\frac{\sqrt{3}d_{\ell m}r^n}{2(\eta_3r_c^2\rho_c)^{1/2}}
        \qquad~~~~~ (\alpha=0)}
    \end{cases}
    ,
\end{align}
leading to the solution
\begin{align}
    \dpi\simeq
    \begin{cases}
        \displaystyle{-\frac{d_{\ell m}r^{n+2}}{(\ell-n-2)(\ell+n+3)(3\alpha)^{1/3}(r_c^2\rho_c)^{2/3}}
        +A_0r^\ell
        \qquad (\alpha\neq 0)}
        \\
        \displaystyle{-\frac{\sqrt{3}d_{\ell m}r^{n+2}}{2(\ell-n-2)(\ell+n+3)(\eta_3r_c^2\rho_c)^{1/2}}
        +A_0r^\ell
        \qquad~~~~~ (\alpha=0)}
    \end{cases}
    ,
\end{align}
where we only present the case of $\ell\neq n+2$.
Having thus obtained the boundary conditions at $r=0$ and at infinity, we can now solve eq.~\eqref{eq:flat-g-pert} with a shooting method.

\begin{figure}
  \begin{minipage}{0.48\columnwidth}
    \centering
    \includegraphics[width=\linewidth]{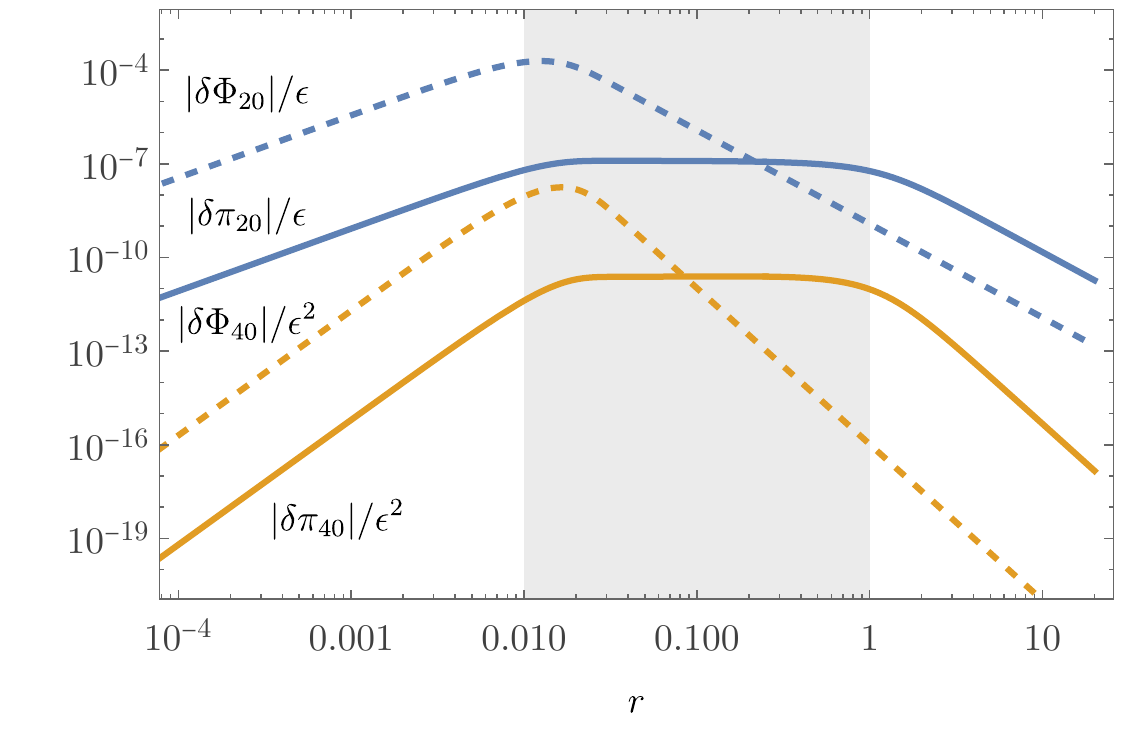}
  \end{minipage}
  \hfill
  \begin{minipage}{0.48\columnwidth}
    \centering
    \includegraphics[width=\linewidth]{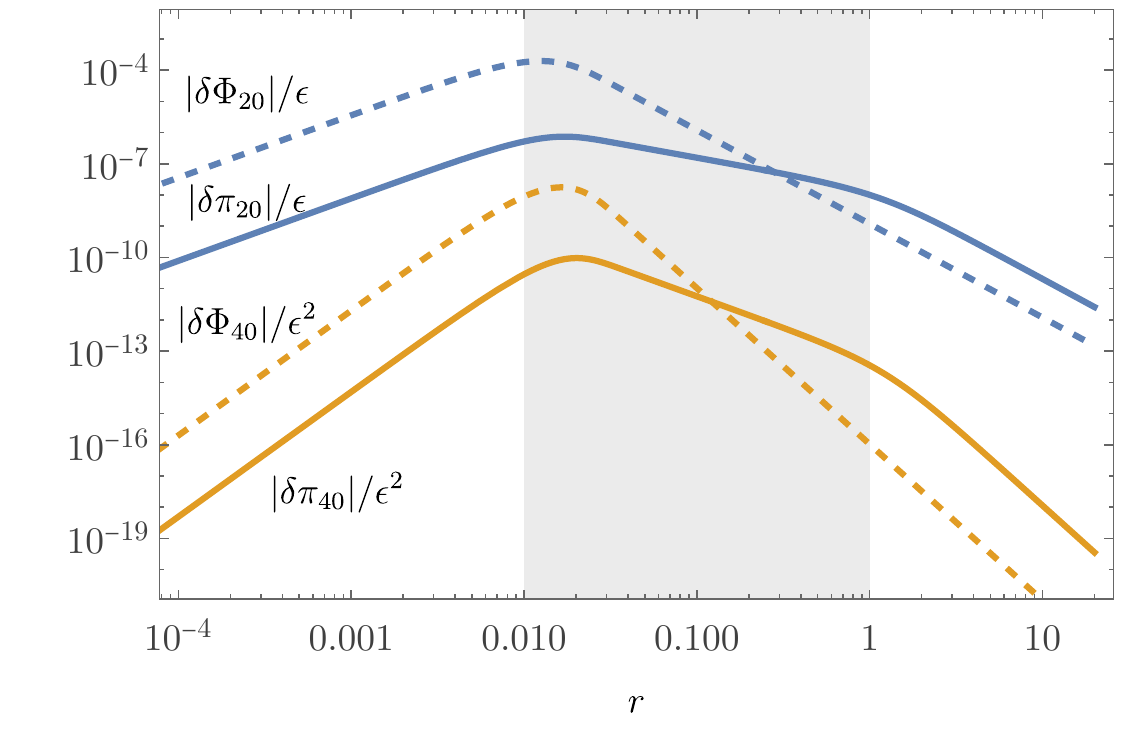}
  \end{minipage}
  \caption{Multipole moments $\dpi$ are compared to those of the standard Newtonian potential $\delta\Phi_{\ell m}$.
  The parameters are given by $\eta_0=\eta_3=1$ and $\alpha=1$ for the left plot and $\eta_0=\eta_3=1$ and $\alpha=0$ for the right plot. The parameters characterizing the density are the same as those used in the main text. The shaded region corresponds to $R_0\le r\le r_{\textrm{V}}$.}
  \label{fig:flat-g}
\end{figure}

We show in figure~\ref{fig:flat-g} the results for the same Gaussian density profile as given in section~\ref{subsec:example}.
The galileon multipole moments $\dpi$ should be compared with the Newtonian counterpart, $\delta\Phi_{\ell m}$, obtained by integrating $r^{-2}\left(r^2\delta\Phi_{\ell m}'\right)'-\ell(\ell+1)\delta\Phi_{\ell m}/r^2=\delta\rho_{\ell m}/2$.
We see that the multipole moments of the fifth force from the galileon are suppressed compared to the Newtonian counterpart for $r\ll r_{\textrm{V}}$, indicating efficient Vainshtein screening in the case of the flat-space galileon.

The behavior of $\dpi$ in the region $R_0\lesssim R\lesssim r_{\textrm{V}}$ is as follows.
In that region, $\pi_0'$ is given by eq.~\eqref{eq:flat-g-soln-Vianshtein}, and hence the homogeneous solution to eq.~\eqref{eq:flat-g-pert} for $x_0\gg 1$ is
\begin{align}
    \dpi=
    \begin{cases}
        C_0+C_1r\qquad\qquad\qquad~~~ (\alpha\neq 0)
        \\ 
        C_2 r^{(\ell+1)/2}+C_3 r^{-\ell/2}
        \qquad (\alpha=0)
    \end{cases}.
\end{align}
Our numerical results imply that the nongrowing part is dominant, $\dpi\simeq C_0$ for $\alpha\neq 0$ and $\dpi\propto r^{-\ell/2}$ for $\alpha=0$.

The results presented here are in sharp contrast to those for the EFT of dark energy with $\alpha_H+2\beta_1\neq 0$ obtained in section~\ref{sec:multip}.

\section{Cylindrically symmetric solution}\label{app:cylindrical}

Let us consider cylindrically symmetric configurations, for which all the quantities depend only on $\varrho:=\sqrt{x^2+y^2}$ and eqs.~\eqref{eq:tildePhi} and~\eqref{eq:reduce-scalar-eq} can be solved easily:
\begin{align}
    \D s^2&=-[1+2\Phi(\varrho)]\D t^2+[1-2\Psi(\varrho)]
    \left(\D\varrho^2+\varrho^2\D\theta^2+\D z^2\right),
    \\ 
    \pi&=\pi(\varrho),
    \\ 
    \rho&=\rho(\varrho).
\end{align}
Equations~\eqref{eq:tildePhi} and~\eqref{eq:reduce-scalar-eq} give, respectively,
\begin{align}
    \tilde y=\frac{\xi}{2}x+\frac{1-\alpha_H-3\beta_1}{2(1+\alpha_H+\beta_1)^2}
    \mathcal{B}+\frac{\alpha_H+2\beta_1}{2(1+\alpha_H+\beta_1)^2}x^2,
    \label{ap-yeq}
\end{align}
and
\begin{align}
    \hat \eta_0 x-\frac{\xi}{2}(1-\alpha_H-3\beta_1)\mathcal{B}
    +\frac{\hat\eta_3}{2}x^2
    -2(\alpha_H+2\beta_1)x\tilde y
    -\frac{(\alpha_H+\beta_1)(1-\alpha_H-3\beta_1)}{1+\alpha_H+\beta_1}
    \frac{(\varrho^2\mathcal{B})'}{\varrho}x=0,
    \label{ap-xeq}
\end{align}
where
\begin{align}
    x:=\frac{r_c^2\pi'}{\varrho},
    \qquad 
    \tilde y:=\frac{r_c^2\tPhi'}{\varrho},
    \qquad 
    \mathcal{B}:=\frac{r_c^2\lambda}{\varrho^2},
\end{align}
with $\lambda'=\varrho\rho$, and a dash denotes differentiation with respect to $\varrho$ in this appendix.
Our goal is to calculate the dimensionless gradient of the gravitational potentials,
\begin{align}
    y&:=\frac{r_c^2\Phi'}{\varrho}=\tilde y-\frac{\alpha_H+\beta_1}{1+\alpha_H+\beta_1}
    x(\varrho x)',
    \\ 
    z&:=\frac{r_c^2\Psi'}{\varrho}=(1+\alpha_H)y+\frac{\eta_2}{4}x+\alpha_H
    x(\varrho x)',
    \label{app-eqz}
\end{align}
to see whether or not they exhibit the expected force law $y,z \propto \mathcal{B}$ in the exterior of the matter distribution (where $\lambda=\lambda_0=$ const) and coincide, $y=z$, within a certain radius (or, more specifically, $\varrho_{\textrm{V}}:=r_c\lambda_0^{1/2}$).

\begin{figure}
  \begin{minipage}{0.48\columnwidth}
    \centering
    \includegraphics[width=\linewidth]{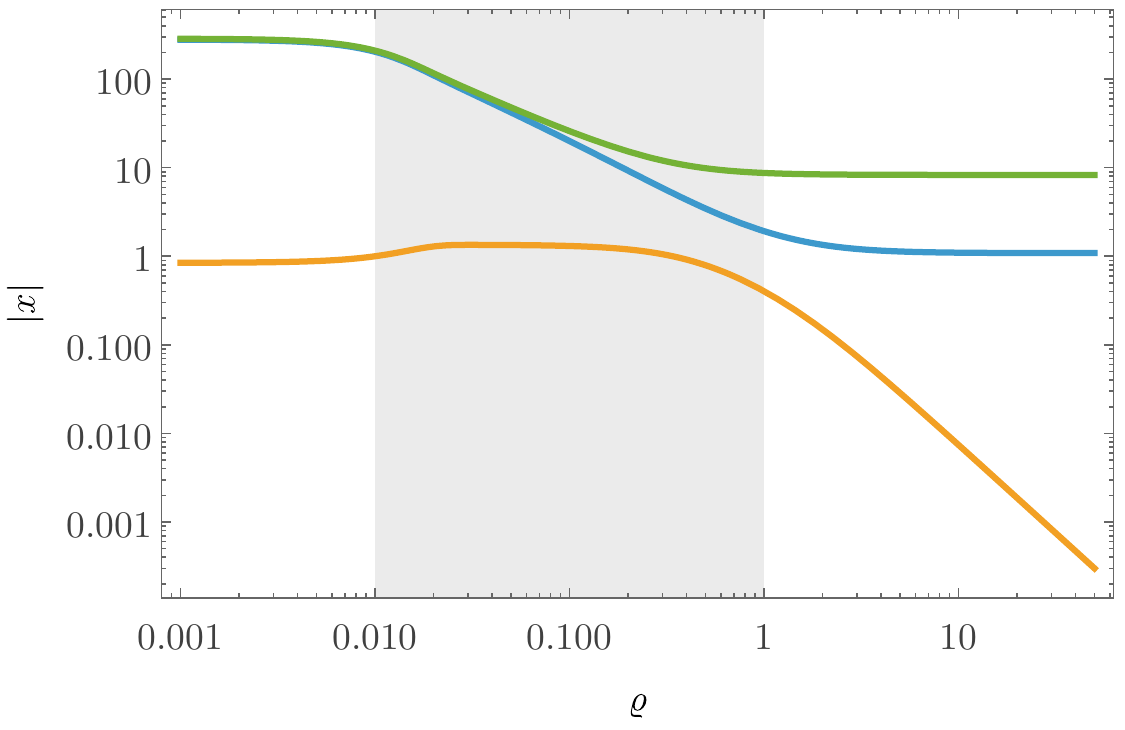}
  \end{minipage}
  \hfill
  \begin{minipage}{0.48\columnwidth}
    \centering
    \includegraphics[width=\linewidth]{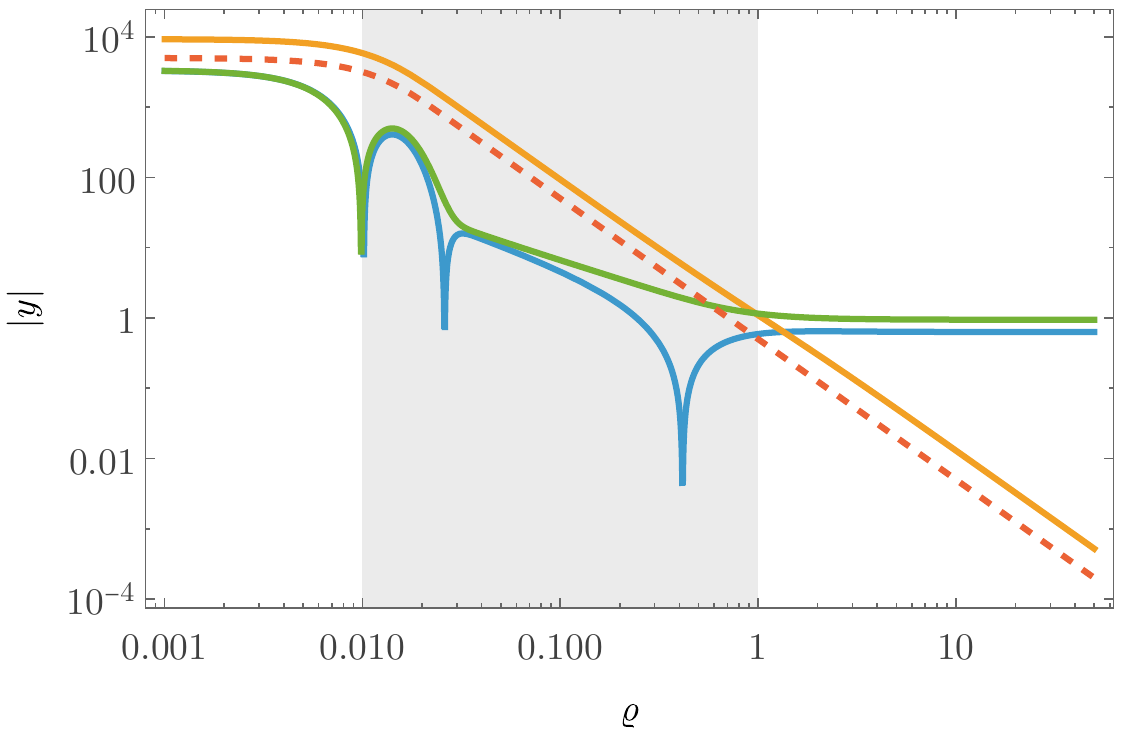}
  \end{minipage}
  \caption{Solutions to eqs.~\eqref{ap-yeq} and~\eqref{ap-xeq} for the density profile~\eqref{ap-density-pro} with $\varrho_{\textrm{V}}=r_c\lambda_0^{1/2}=1$ and $R_0=0.01$.
  For comparison, the solution to $\nabla^2\Phi_{\textrm{N}}=\rho/2$, i.e., $r_c^2\Phi_{\textrm{N}}'/\varrho=\mathcal{B}/2$ is shown in the red dashed line.
  The EFT parameters are given by $\hat\eta_0=\xi=\hat\eta_3=1$, $\alpha_H=0.1$, and $\beta=-0.2$.
  The shaded region corresponds to $R_0\le\varrho\le \varrho_{\textrm{V}}$.}
  \label{fig:cyl}
\end{figure}

Removing $\tilde y$ from eq.~\eqref{ap-xeq} with the help of eq.~\eqref{ap-yeq}, we obtain a cubic equation for $x$, provided that $\alpha_H+2\beta_1\neq 0$, admitting three branches of solutions.
Among them, we look for a solution that is connected to the linear solution
\begin{align}
    x=\frac{\xi}{2\hat\eta_0}(1-\alpha_H-3\beta_1)\mathcal{B}
    \label{ap-linear-soln}
\end{align}
for $\varrho\gg\varrho_{\textrm{V}}$.
In the nonlinear regime where $\mathcal{B}\gg 1$, we have the following three solutions:
\begin{align}
    x&\simeq \pm \sqrt{
        \frac{1-\alpha_H-3\beta_1}{\alpha_H+2\beta_1}
        \left[-\mathcal{B}+\frac{\alpha_H+\beta_1}{\alpha_H+2\beta_1}
        \frac{(\varrho^2\mathcal{B})'}{\varrho}\right]
    }\gg 1,
    \\ 
    x&\simeq -\frac{\xi(1+\alpha_H+\beta_1)^2\mathcal{B}}%
    {2[(\alpha_H+2\beta_1)\mathcal{B}+(\alpha_H+\beta_1)(1+\alpha_H+\eta_1)(\varrho^2\mathcal{B})'/\varrho]}=\mathcal{O}(1).
    \label{ap-3rd-branch}
\end{align}
The first two branches correspond to the solution giving the screened result for spherically symmetric configurations.
In the cylindrically symmetric case, however, it turns out that $y$ depends on the source only through $\lambda'$ and $\lambda''$ in these branches, implying that the expected force law is not recovered.
In the third branch, we have $z=(1+\alpha_H)y\propto\mathcal{B}$ (as seen from eq.~\eqref{app-eqz}), and hence the two potentials do not coincide despite the expected $y\propto \mathcal{B}$ behavior.
Thus, the screening mechanism does not work well.
For the Gaussian profile
\begin{align}
    \rho=\frac{2\lambda_0}{R_0^2}e^{-\varrho^2/R_0^2},
    \qquad 
    \lambda=\lambda_0\left(1-e^{-\varrho^2/R_0^2}\right),
    \label{ap-density-pro}
\end{align}
we show the three branches in figure~\ref{fig:cyl}.
In this case, the third solution~\eqref{ap-3rd-branch} is connected to the linear solution~\eqref{ap-linear-soln}.

In the special case of $\alpha_H+2\beta_1=0$, eqs.~\eqref{ap-yeq} and~\eqref{ap-xeq} reduce respectively to
\begin{align}
    &\tilde y=\frac{\xi}{2}x+\frac{\mathcal{B}}{2(1-\beta_1)},
    \\ 
    &\hat\eta_0 x-\frac{\xi}{2}(1-\beta_1)\mathcal{B}+\frac{\hat\eta_3}{2}x^2
    +\beta_1\frac{(\varrho^2\mathcal{B})'}{\varrho}x=0.
\end{align}
Since we have a quadratic equation for $x$ in this case, there are two branches.
In the exterior of the source, we have $x\propto \mathcal{B}^{1/2}\propto \varrho^{-1}$ for $\mathcal{B}\gg 1$, irrespective of the branch.
Thus, we find that $y=\mathcal{B}/[2(1-\beta_1)]$ and $z=(1-2\beta_1)y$.
Also in this case, there is no screened solution.

We conclude that the fifth force is not screened for cylindrically symmetric configurations in DHOST theories and in the EFT of dark energy with nonvanishing $\alpha_H$ and $\beta_1$.
This kind of shape dependence of the screening mechanism would have an impact on the cosmic large-scale structure~\cite{Falck:2014jwa, Falck:2015rsa, Burrage:2019afs}.

\section{Horndeski with \texorpdfstring{$\alpha_T\neq 0$}{alphaT neq 0}}
\label{app:Horn}

In this appendix, we show that a similar nonstandard behavior of the multipole moments of the gravitational potential is seen within the Horndeski family of scalar-tensor theories if one allows for nonluminal gravitational wave propagation, $\alpha_T\neq 0$.
Setting the ``beyond Horndeski'' EFT parameters to zero but now allowing for $\alpha_T\neq 0$, the effective Lagrangian reads
\begin{align}
        \mathcal{L}_{\textrm{eff}}&=\frac{M^2}{2}\bigl[
                \left(
                        \eta_0\pi+\eta_1\Phi+\eta_2\Psi
                \right)\nabla^2\pi + 4\Psi\nabla^2\Phi -2(1+\alpha_T) \Psi\nabla^2\Psi 
                \notag \\ & \quad 
                +r_c^2\eta_3\mathcal{L}_3^{\textrm{Gal}}
                +r_c^2\left(\alpha_V\Phi
                +\alpha_T\Psi\right)\mathcal{E}_3^{\textrm{Gal}}
                        -r_c^4(\alpha_T+\alpha_V)\mathcal{L}_4^{\textrm{Gal}}
        \bigr]-\Phi \rho.
        \label{effLag-alphaT}
\end{align}
Below in this appendix we absorb the factor $M^2(1+\alpha_V)$ into $\rho$ and write $\rho/M^2(1+\alpha_V)\to \rho$ so that the effective Newton constant is equal to $1/8\pi$.
The field equations are given by
\begin{align}
    \frac{\eta_1}{2}\nabla^2\pi+2\nabla^2\Psi+\frac{r_c^2\alpha_V}{2}\mathcal{E}_3^{\textrm{Gal}}
    -(1+\alpha_V)\rho&=0,
    \label{apd:01}
    \\ 
    \frac{\eta_2}{2}\nabla^2\pi+2\nabla^2\Phi-2(1+\alpha_T)\nabla^2\Psi
    +\frac{r_c^2\alpha_T}{2}\mathcal{E}_3^{\textrm{Gal}}&=0,
    \label{apd:02}
\end{align}
and
\begin{align}
    &
    \eta_0\nabla^2\pi
    +\frac{\eta_1}{2}\nabla^2\Phi+\frac{\eta_2}{2}\nabla^2\Psi
    +\frac{r_c^2\eta_3}{2}\mathcal{E}_3^{\textrm{Gal}}
    +r_c^2\alpha_V\left[\nabla^2\Phi\nabla^2\pi-\Phi_{ij}\pi_{ij}\right]
    \notag \\ & 
    +r_c^2\alpha_T\left[\nabla^2\Psi\nabla^2\pi-\Psi_{ij}\pi_{ij}\right]
    -\frac{r_c^4}{2}(\alpha_T+\alpha_V)\mathcal{E}_4^{\textrm{Gal}}
    =0.\label{apd:03}
\end{align}
From eqs.~\eqref{apd:01} and~\eqref{apd:02} we obtain
\begin{align}
    \nabla^2\Phi&=\frac{1}{2}(1+\alpha_T)(1+\alpha_V)\rho
    +\frac{\xi}{2}\nabla^2\pi 
    -\frac{r_c^2}{4}\left[\alpha_T+(1+\alpha_T)\alpha_V\right]\mathcal{E}_3^{\textrm{Gal}},
    \\ 
    \nabla^2\Psi&=\frac{1}{2}(1+\alpha_V)\rho
    -\frac{\eta_1}{4}\nabla^2\pi 
    -\frac{r_c^2}{4}\alpha_V\mathcal{E}_3^{\textrm{Gal}},
\end{align}
where $\xi:=-[\eta_1(1+\alpha_T)+\eta_2]/2$.
Using these equations one can rewrite eq.~\eqref{apd:03} into the form
\begin{align}
    &
    \hat\eta_0\nabla^2\pi-\frac{\xi}{2}(1+\alpha_V)\rho+\frac{r_c^2\hat\eta_3}{2}
    \mathcal{E}_3^{\textrm{Gal}}+r_c^2\alpha_V\left[\nabla^2\Phi\nabla^2\pi-\Phi_{ij}\pi_{ij}\right]
    \notag \\ & 
    +r_c^2\alpha_T\left[\nabla^2\Psi\nabla^2\pi-\Psi_{ij}\pi_{ij}\right]
    -\frac{r_c^4}{2}(\alpha_T+\alpha_V)\mathcal{E}_4^{\textrm{Gal}}
    =0,
\end{align}
where the explicit expressions for $\hat\eta_0$ and $\hat\eta_3$ are not necessary for the present purpose.

Let us first consider a spherically symmetric solution, $\pi=\pi_0(r)$, $\Phi=\Phi_0(r)$, $\Psi=\Psi_0(r)$, for a spherically symmetric density profile $\rho=\rho_0(r)$.
We introduce the dimensionless variables $x_0:=r_c^2\pi_0'/r$, $y_0=r_c^2\Phi_0'/r$, $z_0:=r_c^2\Psi_0'/r$, and $\mathcal{A}:=r_c^2\mu/r^3$, with $\mu':=r^2\rho_0$.
The field equations then read
\begin{align}
    y_0&=\frac{1}{2}(1+\alpha_T)(1+\alpha_V)\mathcal{A}-\frac{\xi}{2}x_0
    -\frac{1}{2}\left[\alpha_T+(1+\alpha_T)\alpha_V\right]x_0^2,
    \\ 
    z_0&=\frac{1}{2}(1+\alpha_V)\mathcal{A}-\frac{\eta_1}{2}x_0-\frac{\alpha_V}{2} x_0^2,
\end{align}
and 
\begin{align}
    \hat \eta_0x_0-\frac{\xi}{2}(1+\alpha_V)\mathcal{A}+ \hat\eta_3
    x_0^2+2\alpha_Vx_0 y_0+2\alpha_Tx_0z_0-(\alpha_T+\alpha_V)x_0^3=0.
\end{align}
In the nonlinear regime where $\mathcal{A}=(r_{\textrm{V}}/r)^3\gg 1$, the physically interesting solution is given by
$x_0^2\simeq \mathcal{A}$, $y_0=z_0=\mathcal{A}/2$~\cite{Kimura:2011dc}.
Thus, the Vainshtein mechanism is seen to work for spherically symmetric configurations.

Let us next consider small deviations from spherical symmetry as in the main text.
The multipole components obey
\begin{align}
    &\left[
        \frac{1}{r^2}\left(r^2\delta\Phi_{\ell m}'\right)'-\frac{\ell(\ell+1)}{r^2}
        \delta\Phi_{\ell m}
    \right] -\frac{\xi}{2}
    \left[
        \frac{1}{r^2}\left(r^2\dpi'\right)'-\frac{\ell(\ell+1)}{r^2}\dpi
    \right]
    \notag \\ &
    +\frac{1}{2}\left[\alpha_T+(1+\alpha_T)\alpha_V\right]
    \left[
        \frac{2}{r^2}\left(r^2x_0\dpi'\right)'
    -\frac{\ell(\ell+1)}{r^3}\left(r^2x_0\right)'\dpi
    \right]=\frac{1}{2}(1+\alpha_T)(1+\alpha_V)\delta\rho_{\ell m},
    \\ 
    &\left[
        \frac{1}{r^2}\left(r^2\delta\Psi_{\ell m}'\right)'-\frac{\ell(\ell+1)}{r^2}
        \delta\Psi_{\ell m}
    \right] +\frac{\eta_1}{4}
    \left[
        \frac{1}{r^2}\left(r^2\dpi'\right)'-\frac{\ell(\ell+1)}{r^2}\dpi
    \right]
    \notag \\ &
    +\frac{\alpha_V}{2}
    \left[
        \frac{2}{r^2}\left(r^2x_0\dpi'\right)'
    -\frac{\ell(\ell+1)}{r^3}\left(r^2x_0\right)'\dpi
    \right]=\frac{1}{2}(1+\alpha_V)\delta\rho_{\ell m},
\end{align}
and
\begin{align}
    &\hat\eta_0\left[
        \frac{1}{r^2}\left(r^2\dpi'\right)'-\frac{\ell(\ell+1)}{r^2}\dpi
    \right]-\frac{\xi}{2}(1+\alpha_V)\delta\rho_{\ell m}
    \notag \\ &
    +
    \hat\eta_3\left[
        \frac{2}{r^2}\left(r^2x_0\dpi'\right)'
    -\frac{\ell(\ell+1)}{r^3}\left(r^2x_0\right)'\dpi
    \right]
    \notag \\ & 
    +\alpha_V
    \left[
    \frac{2}{r^2}\left(r^2x_0\delta\Phi'_{\ell m}\right)'-\ell(\ell+1)
    \frac{(r^2x_0)'}{r^3}\delta\Phi_{\ell m}
    +\frac{2}{r^2}\left(r^2y_0\dpi'\right)'-\ell(\ell+1)
    \frac{(r^2y_0)'}{r^3}\dpi
    \right]
    \notag \\ & 
    +\alpha_T
    \left[
    \frac{2}{r^2}\left(r^2x_0\delta\Psi'_{\ell m}\right)'-\ell(\ell+1)
    \frac{(r^2x_0)'}{r^3}\delta\Psi_{\ell m}
    +\frac{2}{r^2}\left(r^2z_0\dpi'\right)'-\ell(\ell+1)
    \frac{(r^2z_0)'}{r^3}\dpi
    \right]
    \notag \\ & 
    -\frac{3}{2}(\alpha_T+\alpha_V)
    \left[
        \frac{2}{r^2}\left(
        r^2x_0^2\dpi'
    \right)'-\frac{\ell(\ell+1)}{r^3}\left(r^2x_0^2\right)'\dpi
    \right]=0.
\end{align}
In the nonlinear regime, we may take the limit $x_0^2=2y_0=2z_0=(r_{\textrm{V}}/r)^3\gg 1$, yielding
\begin{align}
    &\left[
        \frac{1}{r^2}\left(r^2\delta\Phi_{\ell m}'\right)'-\frac{\ell(\ell+1)}{r^2}
        \delta\Phi_{\ell m}
    \right] 
    \notag \\ &
    +\frac{1}{2}\left[\alpha_T+(1+\alpha_T)\alpha_V\right]
    \left[
        \frac{2}{r^2}\left(r^2x_0\dpi'\right)'
    -\frac{\ell(\ell+1)}{r^3}\left(r^2x_0\right)'\dpi
    \right]=0,
    \\ 
    &\left[
        \frac{1}{r^2}\left(r^2\delta\Psi_{\ell m}'\right)'-\frac{\ell(\ell+1)}{r^2}
        \delta\Psi_{\ell m}
    \right]
    \notag \\ &
    +\frac{\alpha_V}{2}
    \left[
        \frac{2}{r^2}\left(r^2x_0\dpi'\right)'
    -\frac{\ell(\ell+1)}{r^3}\left(r^2x_0\right)'\dpi
    \right]=0,
\end{align}
and
\begin{align}
    &\alpha_V
    \left[
    \frac{2}{r^2}\left(r^2x_0\delta\Phi'_{\ell m}\right)'-\ell(\ell+1)
    \frac{(r^2x_0)'}{r^3}\delta\Phi_{\ell m}
    +\frac{2}{r^2}\left(r^2y_0\dpi'\right)'-\ell(\ell+1)
    \frac{(r^2y_0)'}{r^3}\dpi
    \right]
    \notag \\ & 
    +\alpha_T
    \left[
    \frac{2}{r^2}\left(r^2x_0\delta\Psi'_{\ell m}\right)'-\ell(\ell+1)
    \frac{(r^2x_0)'}{r^3}\delta\Psi_{\ell m}
    +\frac{2}{r^2}\left(r^2z_0\dpi'\right)'-\ell(\ell+1)
    \frac{(r^2z_0)'}{r^3}\dpi
    \right]
    \notag \\ & 
    -\frac{3}{2}(\alpha_T+\alpha_V)
    \left[
        \frac{2}{r^2}\left(
        r^2x_0^2\dpi'
    \right)'-\frac{\ell(\ell+1)}{r^3}\left(r^2x_0^2\right)'\dpi
    \right]=0,
\end{align}
where we set $\delta\rho_{\ell m}=0$.
We look for the solution in the form of a power law, $\delta\Phi_{\ell m}=A_* r^\gamma$, $\delta\Psi_{\ell m}=B_* r^\gamma$, $\dpi= r^{\gamma+3/2}$, and find the independent solutions with
\begin{align}
    \gamma=-\frac{1}{2}\pm \frac{\sqrt{D_\pm}}{2},
\end{align}
where 
\begin{align}
    D_\pm&=\ell(\ell+1)+\frac{5}{2}\pm\frac{3}{2}
    \sqrt{1-4\kappa_1\ell(\ell+1)
    +4\kappa_2[\ell(\ell+1)]^2},
    \\
    \kappa_1&=\kappa_2-
    \frac{\alpha_V(\alpha_V+2\alpha_T+\alpha_T\alpha_V)}{(1+\alpha_V)(\alpha_T+\alpha_V+\alpha_T\alpha_V)}
    ,
    \\
    \kappa_2&=\frac{\alpha_T+\alpha_V}{(1+\alpha_V)(\alpha_T+\alpha_V+\alpha_T\alpha_V)}.
\end{align}
When $|\alpha_T|\ll 1$ and $|\alpha_V|\ll 1$, the leading terms are the same as those given in eq.~\eqref{eq:powers-gamma}.
This result shows that the main conclusion of this paper is not restricted within the DHOST family of theories.

\bibliography{refs}
\bibliographystyle{JHEP}
\end{document}